\documentclass[%
 reprint,
superscriptaddress,
 amsmath,amssymb,
 aps,
 pra,
]{revtex4-2}

\usepackage{graphicx}
\usepackage{dcolumn}
\usepackage{bm}
\usepackage{hyperref}
\usepackage[mathlines]{lineno}
\usepackage{lipsum}
\usepackage{xcolor}
\usepackage{siunitx}
\usepackage{braket} 
\usepackage{placeins} 
\usepackage{comment}

\usepackage{tocloft}
\usepackage{array}
\newcolumntype{R}[1]{>{\raggedleft\arraybackslash}p{#1}}

\begin{document}


\preprint{APS/123-QED} 

\title{A modular quantum gas platform}

\author{Tobias~Hammel}
\thanks{These authors contributed equally to this work.\\
Corresponding authors: hammel@physi.uni-heidelberg.de, mkaiser@physi.uni-heidelberg.de.}

\author{Maximilian~Kaiser}
\thanks{These authors contributed equally to this work.\\
Corresponding authors: hammel@physi.uni-heidelberg.de, mkaiser@physi.uni-heidelberg.de.}

\author{Daniel~Dux}
\affiliation{Physikalisches Institut der Universität Heidelberg, Im Neuenheimer Feld 226, 69120 Heidelberg, Germany}

\author{Philipp~M.~Preiss}%
\affiliation{Max Planck Institute of Quantum Optics, Hans-Kopfermann-Str. 1, 85748 Garching, Germany}

\author{Matthias~Weidemüller}

\author{Selim~Jochim}
\affiliation{Physikalisches Institut der Universität Heidelberg, Im Neuenheimer Feld 226, 69120 Heidelberg,
Germany}

\date{\today}


\begin{abstract}
    We report on the development of a modular platform for programmable quantum simulation with atomic quantum gases. The platform is centered around exchangeable optical modules with versatile functionalities. The performance of each module is disentangled from all others, enabling individual validation and maintenance of its outputs. The relative spatial positioning of the modules with respect to the position of the atomic sample is set by a global reference frame. In this way, the platform simplifies re-configuration and upgrading of existing setups and accelerates the design of new machines in a time- and cost-efficient manner. Furthermore, it facilitates collaboration among different experimental groups. This standardized hardware design framework, which we call \textit{Heidelberg Quantum Architecture}, paves the way towards a new generation of on-demand and highly adaptable quantum simulation experiments.
\end{abstract}


\maketitle



\section{\label{sec:introduction}Introduction} 

Since the first creation of ultracold quantum gases around 30 years ago, the field evolved from studying the basic properties of quantum degenerate gases towards using them to investigate systems of higher complexity. The experiments require more and more sophisticated functionalities, which led to the development and construction of hundreds of unique quantum gas experiments \cite{UANews}. These employ ultracold neutral atoms of bosonic or fermionic nature \cite{Bloch2012, Bloch2017, Navon2021, Gross2021, Kaufman2021, Vale2021}, dipolar atoms \cite{Norcia2021} and atomic mixtures \cite{Baroni2024}, cold molecules \cite{Kaufman2021, Cornish2024}, Rydberg atoms \cite{Browaeys2020, Jones_2017} and ions \cite{Blatt2012}. To manage the increasing complexity of such machines, it is beneficial to identify parts that can be modularized into smaller, more comprehensible subsystems \cite{Baccarini1996, Parnas72}.

Our approach has three essential features: First, the modularization of optical setups and the introduction of interfaces for tailored and intuitive realization of atom-light interaction. 

Second, the integration of all interfaces into a single compact monolithic structure to enhance stability and positioning accuracy.

And third, the introduction of a global coordinate system and the ability to measure the position of every component within this frame of reference.

The concepts of this platform, which we named \textit{Heidelberg Quantum Architecture}, can be utilized for studies of systems not only in the area of neutral atoms, but can also be extended to the areas of atomic mixtures, Rydberg atoms, molecules, and ions.

In this paper, we first present the concept of our approach to building experimental hardware in Section~\ref{sec:rational} and elaborate on the technical details of a realization of an ultracold Fermi gas experiment utilizing $^6$Li based on this approach in Section~\ref{sec:CompMod}, including the development of a high-flux atom source. In Section~\ref{sec:Example}, we illustrate the broad applicability of the concepts by explaining the workflow for configuring the platform to realize an experiment with unique functionalities. In Section~\ref{sec:Outlook} we conclude by discussing the advantages of modularization in the context of quantum simulation experiments.

\section{\label{sec:rational}Rationale of the platform architecture} 

\begin{figure}[h]
\includegraphics{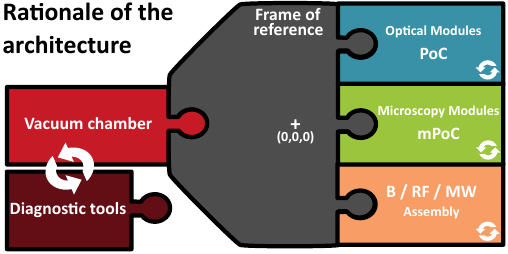}
\caption{\label{fig:rationale} Schematic of the modular concept of this experimental platform. It revolves around a central structure providing a common frame of reference and interfacing points for all modules. The vacuum chamber is retractable from the frame of reference to place diagnostic tools at the position of the atoms. In this way, the optical modules can be externally developed, tested, and implemented in the platform on fast timescales. PoC: Pieces of Cake, mPoC: microscopy Pieces of Cake, B: magnetic field, RF: radio frequency, MW: microwave.}
\end{figure}

\begin{figure*}
\includegraphics{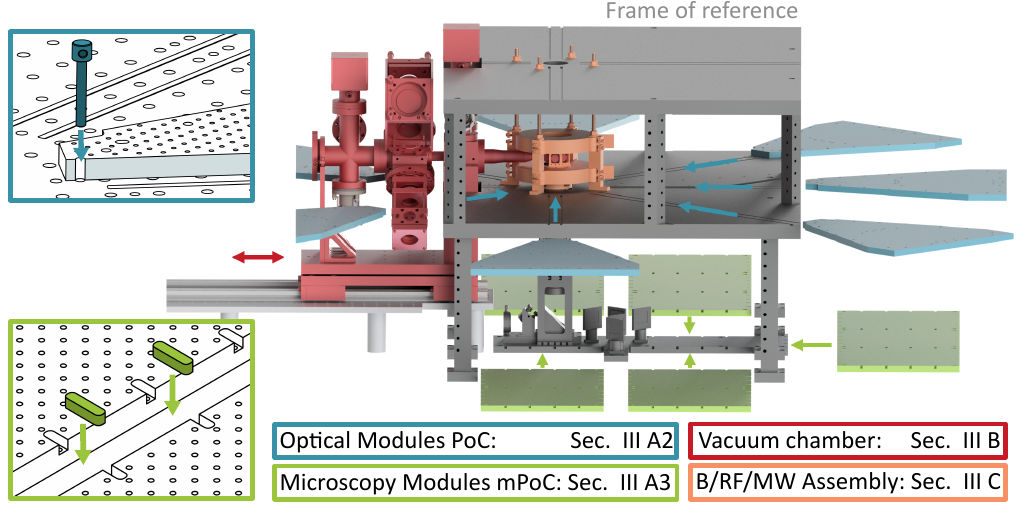}
\caption{\label{fig:full-cad}A simplified CAD model of the freely configurable \textit{Heidelberg Quantum Architecture}. The color coding of the individual parts matches their equivalents in Figure~\ref{fig:rationale}. The design is centered around a common mechanical structure (shown in gray) which serves as a frame of reference for all modules. The vacuum system with an octagonal glass cell as the main science chamber (shown in red) is mounted on rails and can be repeatably inserted/removed from the origin of the frame of reference. This origin, passively overlapping with the focal plane of the high-NA objective, is surrounded by an assembly of magnetic field coils (shown in orange). Optical access to the center is provided via the optical modules. These exchangeable modules are passively aligned to the center of the interface with precisions on the order of a few tens of microns. The working principles of mechanical referencing for the modules are shown in the insets. Modules perpendicular to the optical axis of the objective (shown in blue) are referenced with stainless steel pins and modules for propagation through the high-NA objective (shown in green) are placed using stainless steel fitting keys.}
\end{figure*}

To illustrate the basic concept of our approach, we consider an experiment to image single atoms in an ultracold cloud of optically trapped atoms. Such an experiment requires an ultrahigh-vacuum (UHV) chamber that includes an atom source, several sets of near-resonant laser fields for cooling, trapping, and diagnostics, as well as a microscope objective with a large numerical aperture (NA) to resolve the fluorescence of single atoms \cite{Bergschneider_2018}. As schematically shown in Figure~\ref{fig:rationale}, the rationale behind our platform architecture is to disentangle the entire setup into segments, called modules, each being devoted to a specific functional task (cooling light, optical trapping, magnetic fields, imaging, diagnostic tools, etc.). A frame of reference (gray) defines the mechanical connection of modules with distinct functionalities, such as optical (blue and green) or magnetic (orange) ones to the platform. In our example, the center of this reference frame is located close to the final position of all traps and hence the final atom position.

Due to the modularity, each individual module can be tested and calibrated independently before attaching it to the central interface. For instance, an optical module could be a breadboard with optical elements, which has the task of generating a particular light field distribution interacting with the cold atom cloud in a well-defined plane inside the experimental chamber. This light field distribution can be precisely pre-aligned by mapping it onto an optics characterization device outside of the experimental platform.

Such a modular approach takes advantage of the fact that, even for complex quantum gas experiments, tolerances for initial alignment of light or magnetic fields are on the same order of magnitude as tolerances achieved by CNC machines.

By referencing the optics and possibly other modules to a common frame of reference, each module can be pre-adjusted at an external test bench and then precisely integrated mechanically into the platform, thus preserving the absolute orientation to all other components of the setup. Figure~\ref{fig:full-cad} shows a CAD model of the mechanical realization of this concept in our experimental platform. The color coding follows the one in Figure~\ref{fig:rationale} to identify the different modules.

The common frame of reference in our platform is provided by a custom machined board with precision board holes (see Section~\ref{subsubsec:PoCs}). We can position the horizontal optical modules (light blue in Figure~\ref{fig:rationale} and Figure~\ref{fig:full-cad}) with a precision of typically a few tens of microns with respect to the origin of the frame of reference. We achieve this by utilizing precision positioning pins, which are visualized in the blue inset of Figure~\ref{fig:full-cad} and will be discussed in detail in Section~\ref{subsubsec:PoCs}. Due to their shape and the huge facilitation they provide in everyday work, we call them "Pieces of Cake" (PoC).

In terms of sensitivity to misalignment, the most demanding piece of optics is the central high-NA objective. For all optics using the same high-NA objective, for example, to perform high-resolution manipulation, trapping, or imaging, we develop an adapted modularization strategy. For details, we refer to Section~\ref{subsubsec:high NA}. To summarize the idea, an appropriate optical setup introducing a magnification into the optical path can transform position requirements of tens of microns in the atom plane to hundreds of microns or even millimeters in the image plane. The breadboards used for this purpose, shown in light green in Figure~\ref{fig:full-cad}, are called microscopy Pieces of Cake (mPoC).

These two types of optical modules (PoCs and mPOCs) form the core of this modularization concept for optics arrangements in the context of quantum gas experiments. To interface horizontal and vertical optical modules, a precise pre-characterization of the objective and its optical axis is required, as it allows us to position the objective such that its field of view (FOV) is centered around the origin of the global coordinate system given by the frame of reference. Using the same positioning mechanism as for the PoCs, the objective is mounted onto the frame of reference without adjustable degrees of freedom. The angular adjustment of the objective with the vacuum window is critical, as tilting the optical viewport introduces aberrations for high-NA objectives. Precise alignment of this degree of freedom is ensured by tilting the whole vacuum system using spacers, as outlined in Section~\ref{subsec:Vacuum}.

The vacuum system, including the cold atom source, main science chamber and pumping units, is disconnected from the frame of reference, but can readily be moved in and out using a rail system (see Figure~\ref{fig:full-cad}). The absolute positioning of the science chamber compared to the frame of reference is not critical and is defined by placing a stop to the travel range of the rail system.  The vacuum chamber can be moved in and out of the frame of reference providing access to the origin of the reference coordinate system, where the atoms are approximately located, allowing one to place diagnostic tools like cameras or magnetic field sensors. This feature of interchangeability between the vacuum chamber and diagnostic tools is illustrated in Figure~\ref{fig:rationale}.

In addition, connected to the frame of reference are magnetic field and RF coils shown in orange in Figure~\ref{fig:full-cad}. These can in principle be removed or added to the platform, but are not designed for frequent exchange. Technical details of the magnetic field assemblies can be found in Section~\ref{subsec:MagFields}.

A combination of multiple of these modules defines the functionality of the experiment, and in the following will be called "configuration" of the platform. As modules can easily be exchanged, this makes the platform highly versatile.

\section{\label{sec:CompMod}Components and Modules} 

In the following, we are going to describe the technical realization of the modules and assemblies in this quantum gas platform. The design of these we discuss for our specific configuration of the platform, namely a quantum simulation experiment using degenerate gases of fermionic $^6$Li. 

\subsection{\label{subsec:modularity}Modular Experimental Interface} 

A fundamental aspect of achieving modularity for quantum gas platforms is to define a common frame of reference, providing suitable interfacing points between the exchangeable modules and the permanent framework of the apparatus. Chosen interfaces must be tailored to the dependencies, tolerances, and functionalities of the involved subsystems. Thus, depending on the requirements of the subsystems, different types of interfaces are possible to connect modules to the permanent framework.

A fundamental requirement for the modularization of optical setups is to disentangle the light fields generated by the modules, particularly decoupling their alignment and testing processes. We achieve this by aligning light fields not relative to one another (without knowledge of the absolute positioning in space), but with respect to a monolithic structure that serves as an absolute reference for all setups. As a result, our approach moves the frame of reference from an optical definition to a mechanical one.

Before describing the technical implementation of the two realizations of optical modules in this platform (PoCs and mPoCs) we present the underlying concept.

\subsubsection{\label{subsubsec: Optics theory} Introduction to optics modularization}

\begin{figure}[t]
\includegraphics[width=0.48\textwidth]{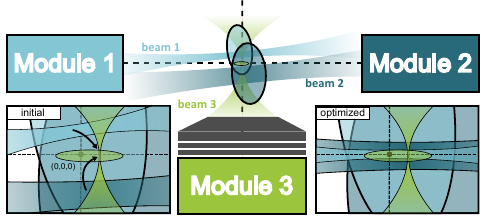}
\caption{\label{fig:Optics theory FOVs 2}A schematic depiction of the passive interfacing concept for the light fields of different optical modules. Each module has an operative region (OR), here depicted as ellipses, in which the light fields fulfill the design parameters. To make them able to work simultaneously all regions need to have a common overlap. Here, as the green OR is fully contained in the ORs of module 1 and 2 the green ellipse also represents the common overlap region of all modules. The actual light fields (labeled beam 1-3), which should perform together, have to be positioned inside this overlap region. In the lower left sketch, a zoom-in of the region around the origin of the frame of reference is shown, initially after placing the two PoCs and the mPoC into the platform. In this example, beam 1 and beam 2 are not inside the overlap region and require fine alignment. In the lower right sketch this fine alignment has been performed. We emphasize here that the ORs of the modules stay constant in position during this process and only the light fields are moved. Placing modules such that there is an initial overlap region of all modules is what we refer to as being passively aligned.}
\end{figure}

The main goal for any optical module integrated into this platform is to be able to generate a well-defined light field (position and angular distribution) at a position where it can intersect with the light fields of all other modules and ultimately the atoms. We refer to the region in space, in which the module can generate the light field it was developed for, as the "operative region" (OR). In this region, for example, the focus of a light beam generating an optical dipole trap can be shifted while retaining full optical performance (shape, trap frequency, etc.). The shape of this OR depends on the functionality of the module as well as the specific way of implementation. It can be limited by, e.g., the field of view of the optics used, the tunability of degrees of freedom or by clipping. The dimensions of such an OR can range from a few wavelengths up to several millimeters. 

Hence, we can define for any optical setup a three-dimensional volume with a well-defined position and size in which the light field produced by the setup satisfies the desired functionality. In Figure~\ref{fig:Optics theory FOVs 2} the ORs of different modules are depicted as ellipses, with a width perpendicular and a depth parallel to the axis of propagation. From this definition, it follows directly that two modules with two separate ORs can only perform simultaneously when the OR volumes overlap. Hence, the challenge to achieve modularization is to place the modules in such a way that the ORs of all modules have a sufficient overlap.

As shown in Figure~\ref{fig:Optics theory FOVs 2} this concept works to overlap ORs along the same axis, on perpendicular axes, or crossing under any angle. The modules presented later are implemented into a common frame of reference with a well-defined origin (0,0,0). They are designed such that this origin is well within the overlap of all ORs. This concept allows us to determine the required position and angle tolerances to achieve passive interfacing of a given set of modules. We estimate these tolerances in Appendix~\ref{App: subsec: Optics theory}.

In the following, this concept will be used to describe the technical realization of the different optical modules in Section~\ref{subsubsec:PoCs} and \ref{subsubsec:high NA}.

\subsubsection{\label{subsubsec:PoCs} Horizontal optics - Pieces of Cake (PoCs)} 

\begin{figure*}[t]
\includegraphics{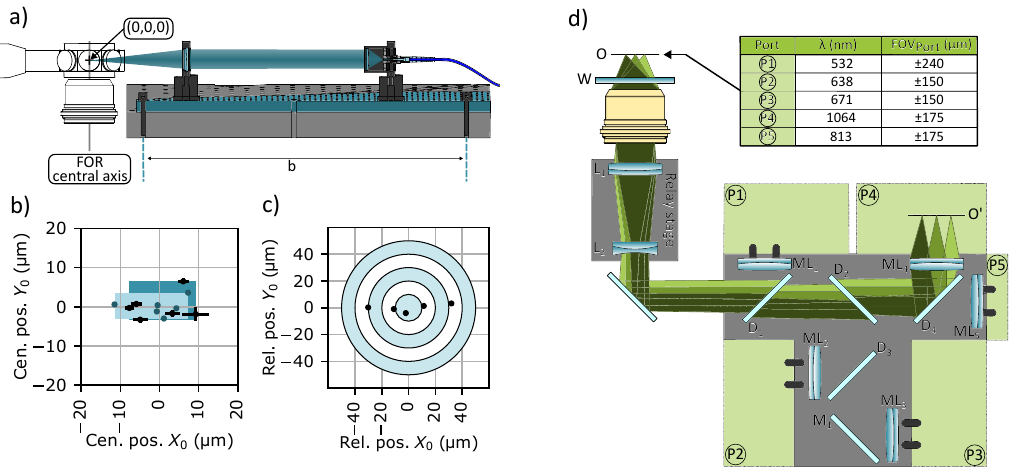}
\caption{\label{fig:modular_tech} \textbf{a)} Modularity concept of the PoCs. Positioning is conducted with two stainless steel pins which set the pointing with respect to the mechanical frame of reference. Positioning and alignment of the optical components on the breadboard are performed externally on a separate test bench. \textbf{b)} The repeatability of placing the PoCs on the mechanical frame of reference is measured on the test bench by sending a collimated beam onto a Shack-Hartman-wavefront sensor in the atom plane and tracing the angle and centroid position when repeatably removing and inserting the PoC. Black and blue data points correspond to two independent PoCs, their respective mean centroid position is set as the reference. We find centroid fluctuations below $\pm$\SI{10}{\micro\meter} in horizontal and below $\pm$\SI{5}{\micro\meter} in vertical direction visualized by the colored rectangles.The dark blue box corresponds to the black points, the light blue box to the blue data points. \textbf{c)} Transferability between the test bench and the experiment is measured by placing in-situ cameras on both the test bench as well as in the atom plane in the experiment and transferring a PoC, producing a collimated beam, multiple times between the interfaces and bolting them down with \SI{1.4}{\newton\meter}. Measuring the relative distance between the centroid positions on the test bench and the atom plane allows us to determine a potential offset in the positioning of the in-situ cameras (see Appendix~\ref{App: subsec: mPoCs}). We find fluctuations of the measured relative distance of less than $\pm$\SI{30}{\micro\meter} in horizontal and below $\pm$\SI{10}{\micro\meter} in the vertical direction. The rings serve as a guide to the eye. \textbf{d)} Modularity concept of the high-NA path and the lower distribution panel. The collimated beam from the rear aperture of the objective is first de-magnified by 3x using a two-lens telescope serving as a relay stage before the beam path gets spectrally split by several dichroic mirrors $D$ and focused by individual, exchangeable module lenses $ML$ onto each port for the high-NA modules P1-P5. This creates an image plane $O'$ of the atom plane $O$ on each module magnified by a variable factor depending on the chosen module lens. The dark rounded rectangles visualize the position of the fitting keys shown in Figure~\ref{fig:full-cad} and hence the mechanical connection points for each port. The table shows the calculated FOV, in which light can propagate unclipped from the atom plane $O$ to the module plane $O'$. This is limited by the aperture of the dichroic mirrors.}
\end{figure*}

Modularization requires making design decisions that work for all possible configurations of the modules intended. In our context, this means that we choose the geometry of the modules to match the optical axes of the vacuum chamber. This results in triangularly shaped optical modules assembled in an octagonal fashion. Using state-of-the-art CNC machining techniques with readily achievable precisions in the \SI{10}{\micro\meter} regime \cite{CNC}, we manufactured a table-like monolithic aluminum framework to serve as a frame of reference in the apparatus, to which all these modules are connected (see Figure~\ref{fig:full-cad} and \ref{fig:modular_tech}). 

The triangularly shaped module breadboards (PoCs) can be positioned on the interface board using two stainless steel alignment pins with a diameter of \SI{4.98}{\milli\meter} that fit into precision-drilled holes with \SI{5.00}{\milli\meter} in diameter (before anodization of the modules). The mechanism is depicted in the inset of Figure~\ref{fig:full-cad} and in Figure~\ref{fig:modular_tech}a. This procedure is simple to use, yet very precise, yielding precisions of around \SI{\pm10}{\micro\meter}. A measurement characterizing the precision is shown in Figure~\ref{fig:modular_tech}b, where a PoC has been repeatedly removed and re-installed using the pin alignment procedure.

The principle of this alignment procedure is broadly applicable for positioning various components relative to one another, as detailed in Section~\ref{subsubsec:high NA}.

In our platform, the PoCs serve as the baseplates of the optical modules. Each module has a predefined functionality that can be pre-aligned and tested individually. This is done on a test bench further described in Appendix~\ref{App: subsec: PoCs}. The ability to build up a PoC on the test bench and place it into the experiment is called transferability. The transferability of a PoC from the test bench to the experiment is shown in Figure~\ref{fig:modular_tech}c). The precision of placing diagnostic tools in the frame of reference is demonstrated for the in-situ cameras and presented in Appendix~\ref{App: subsec: PoCs} and is on the order of $\pm$\SI{20}{\micro\meter}.

This realization of the horizontal optical modules fulfills all criteria described earlier in Section~\ref{subsubsec: Optics theory} and Appendix~\ref{App: subsec: Optics theory} to yield passive interfacing: Optical access of the vacuum chamber in the horizontal direction is limited to 0.3\,NA. Optical components operating in this regime typically have ORs of several hundred micrometers up to millimeters, which is much larger than the positioning precision achieved here. In this way, the operative regions of the different horizontal modules overlap passively, when positioning the PoCs using the described pin alignment procedure.

\subsubsection{\label{subsubsec:high NA} High-NA optics - Microscopy PoCs (mPoCs)} 

Modern quantum simulation experiments require resolution at the limit given by the optical wavelength, utilizing high-NA optics. For different functionalities, possibly at different wavelengths, like optical tweezers and high-resolution imaging, the relative alignment is facilitated by using one shared high-resolution objective for all light fields.

This poses a challenge in the context of modularization as functionalities of different modules can become entangled.

Our solution for this challenge features an additional optical assembly between the atom plane and the modules. This assembly has the task of generating an image of the atom plane for each wavelength. This is shown in Figure~\ref{fig:modular_tech}, where on each port P1 to P5 an image $O'$ of the atom plane $O$ is generated. Hence, the functionalities of this assembly have to include spectral separation of the light fields, allocation of mechanical connection points for the modules, and the implementation of a well-defined transfer function from the modules to the atoms.

To exemplify the technical challenges of realizing such an optical setup, we describe the optical path of one particular module from the image plane to the atoms. For simplicity, one can consider a single tweezer at \SI{1064}{\nano\meter}, which should be freely movable inside the FOV of the objective. This beam path is visualized in Figure~\ref{fig:modular_tech}d). Starting from the atom plane $O$ with a diffraction-limited tweezer the first critical component is the alignment of the objective and the vacuum window $W$. The optical axis of the objective has to be oriented orthogonal to the window surface to support diffraction-limited performance. The tolerances of misalignment for this objective (NA=0.66) is approximately \SI{1}{\milli\radian}. The angle of the vacuum window is adjsuted by tilting the miniaturized vacuum setup, as further described in Appendix~\ref{App: subsec: mPoCs} and \ref{App: subsec: Vacuum design}. This enables us to mount the objective without any degrees of freedom to the frame of reference using the same pin alignment procedure as for the PoCs (see Appendix~\ref{App: subsec: mPoCs}). Thus, the absolute position of the objective is fixed with respect to the horizontal optics, which is important for a passive overlapping of the ORs. Furthermore, due to the monolithic connection of the different optical modules we expect increased relative stability of the generated light fields against disturbances. 

Continuing along the beam path, the light field exits the rear aperture of the objective with an angle depending on the position inside the FOV, here sketched as different shadings of green and ranging up to $\pm$\SI{5}{\milli\radian}. This imposes requirements on the optics used behind the high-NA objective. As we ultimately need to separate different wavelengths and provide connection points for the modules, the beam has to travel for some distance to get to a region, where the assembly can be mounted. In this miniaturized example this distance is below \SI{1}{\meter} for all wavelengths and is limited by the distance between the horizontal plane of PoCs and objective and the optical table. However, this already poses a challenge as the rear aperture of the objective has a size of 1" and the propagation of the beam increases the required size of optics by another $\SI{10}{\milli\radian}\times\SI{1}{\meter}=\SI{10}{\milli\meter}$. Including another factor of $\sqrt{2}$ for the size of mirrors at an angle of \SI{45}{\degree} the use of 3" optics would be required to enable the use of the full FOV of the objective.

Such optical components are often custom-made and more challenging to manufacture in terms of optical properties like flatness. In addition, the options of lenses at such a size become fewer and are limited to larger focal lengths, which hinders the miniaturization of the modules.
As a solution to this challenge, we introduced an additional relay telescope, composed of a \SI{300}{\milli\meter} focal length achromat (Edmund Optics Part-No. 49-378) and a \SI{-100}{\milli\meter} focal length achromat (Thorlabs Part-No. ACN254-100-AB), into the beam path. This de-magnifies the beam by a factor of 3 and relays some of the propagation distance towards the separation optics. The choice of the telescope lenses is a result of extensive Zemax simulations, simulating which telescope design composed of "commercial-of-the-shelf" lenses leads to the least waveform deformations in our setup while still being mountable to the frame of reference. A summary of the design criteria we used for development is given in Appendix~\ref{App: subsec: mPoCs}.

The de-magnified, collimated beam is subsequently spectrally split by a total of four "commercial-off-the-shelf" dichroic mirrors $D_{1},...,D_{4}$~\cite{Dichroics} on a distribution board providing the connection points for the modules (mPoCs), shown schematically in Figure~\ref{fig:modular_tech}d. The table in Figure~\ref{fig:modular_tech}d presents the design wavelengths of the different ports as well as the theoretically achievable FOV in the atom plane. This can in principle be extended to even more wavelengths if required.

To realize diffraction-limited performance for all ports it is essential to have a way of mounting the dichroics without bending them, as this introduces significant astigmatism into the beam path. A suitable mount design as well as the use of slow-curing glue has been found essential in our case and is presented in Appendix~\ref{App: subsec: mPoCs}. Then, for each port (P1 to P5) an exchangeable module lens $ML_{i}$ focuses the collimated light, split onto this port by the dichroic, creating an image of the atom plane $O$ in an intermediate plane $O'$. The magnification in this image plane $O_{i}'$ can be tuned by the choice of the corresponding module lens and is typically within the range of 10\,-\,50\,x (see Appendix~\ref{App: subsec: Optics theory}).

This optical assembly enables modularization of the high-NA optics of different wavelengths. On the one hand, the magnification reduces the positional placement requirements for the modules as discussed in Appendix~\ref{App: subsec: Optics theory} linearly with magnification. In this way, positioning requirements on the order of tens of µm can easily be enlarged to hundreds of µm or even millimeters making placement of high-NA modules easy. On the other hand, the image planes $O_{i}'$ allow for external development and testing of the modules. Given the magnification in the range of 10\,-\,50\,x, the light field distribution in this plane can be monitored by a camera with standard pixel size. Implemented modules can furthermore be continuously monitored by splitting of a copy of the image plane onto a monitoring camera.

The mPoCs themselves are breadboards interfaced via the edge of the distribution board and locked using fitting keys (see Figure~\ref{fig:full-cad}). This provides repeatable positioning of around \SI{50}{\micro\meter}. Combined with the de-magnification of the distribution system this ensures passive positioning of the operative region of all mPoCs with respect to the frame of reference. 

An important feature of this assembly is the fact that all tuning degrees of freedom for the light fields are contained in the modules, while the transfer function between the image planes and the atoms stays constant. This is done by connecting all fixed optics (objective, relay telescope, dichroic mirrors, and module lenses) with sufficient precision to the frame of reference. A detailed discussion of the mechanical requirements and solutions is provided in Appendix~\ref{App: subsec: Optics theory} and \ref{App: subsec: mPoCs}.

In the following, we describe the remaining assemblies in Figure~\ref{fig:full-cad}  utilized in this configuration of the platform, namely the vacuum chamber including the atom source and all magnetic field assemblies. These are specifically tailored for experiments using fermionic $^6$Li, but can easily be adapted for use with other atomic species. 

\begin{figure*}[t]
\includegraphics{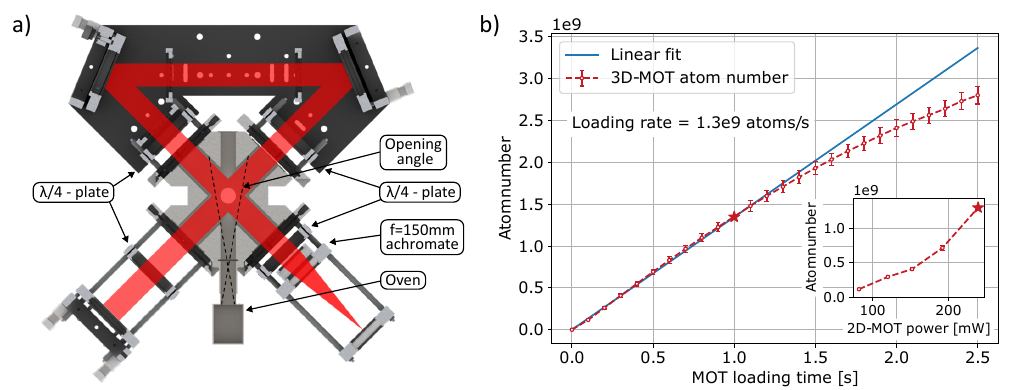}
\caption{\label{fig: MOT_chamber}a) The 2D-MOT chamber with attached optics and oven. The 2D-MOT cooler and repumper beam enter the chamber from the same fiber and for optimal usage of power the beams are assembled in a bowtie. The oven is attached to the bottom CF16 flange of the chamber and the connection neck is designed such that there is no direct line of sight (dashed black lines) of the oven with the viewports. Above the oven there is another CF16 flange that can be used for a Zeeman slower beam. b) 3D-MOT atom numbers for different loading times measured at \SI{350}{\degreeCelsius} oven temperature and the optimized parameters for cooler, repumper and push beam. In blue a linear fit to the non-saturated starting slope is shown yielding loading rates of up to \num{1.3e9} atoms/s.}
\end{figure*}
    
\subsection{\label{subsec:Vacuum}2D-MOT and vacuum} 

An integral part of any quantum gas experiment is a high-flux atom source for the generation of large, degenerate samples of atoms in a short time while supporting an ultra-high vacuum in the science chamber. Here, we present a $^6$Li 2D-MOT design with loading rates of \num{1.3e9} atoms/s at an oven temperature of \SI{350}{\degreeCelsius}, implemented into a vacuum assembly with vacuum lifetimes of $>$\SI{1000}{\second} in the science cell.

There are multiple realizations of 2D-MOTs working as a reliable atom source for different atom species in existing experiments \cite{Chomaz23, Jend22,Nosske_2017,Dorscher_2013,Lamporesi_2013, Tiecke_2009}. For this new design, we use the known scaling behavior of the loading rates with various choosable parameters to increase the capabilities of our cold atom source. Following \cite{Lamporesi_2013,Tiecke_2009}, the loading rates can be scaled up by enlarging the 2D-MOT beams and increasing their detunings from resonance, while providing enough power to be above the saturation intensity of the cooling transition. Following this scaling argumentation, we enlarge the 2D-MOT cooling beams to a diameter of \SI{30}{\milli\meter} and use a single beam bow-tie configuration for optimal power usage (displayed in Figure~\ref{fig: MOT_chamber}a). A strong laser source with \SI{1}{\watt} of nominal output power at \SI{671}{\nano\meter} as well as an optimized oven design (described in Appendix~\ref{App: subsec: Laser Distr.} and \ref{App: subsec: Oven}) further contribute to scaling up the loading rates.

This improved 2D-MOT design is integrated into a vacuum system following the goals of the rationale defined in Section~\ref{sec:rational}. This includes the possibility of extracting the vacuum system from the optical surrounding to place diagnostic tools in the atom plane (see Figure~\ref{fig:rationale}) as well as adjusting the angle of the vacuum window with respect to the high-NA objective. For this, the vacuum system is mounted on a rail system (inspired by \cite{Jend22,Endres_2016}) such that it can easily be moved in and out of the optical surroundings described in Section~\ref{subsec:modularity}. In between the rails and the mounting plate of the vacuum assembly, three stacks of spacers are placed with variable height. By changing the height of individual stacks of spacers we can tilt the vacuum system in steps of \SI{100}{\micro\radian}. To match the angle of the upper viewport of the glass cell to the frame of reference placed an accelerometer with a sensitivity of \SI{100}{\micro\radian} on the glass cell. By previously measuring the orientation of the table one can align both (for details see Appendix~\ref{App: subsec: Oven}). The combination of rail system and spacers proved to be stable and repeatable on the scales relevant for us over the last two years.

A critical design consideration for any atom source has to be the ability to ensure ultrahigh vaccum conditions in the science chamber required for long trapping lifetimes. Here, the simplification of the vacuum, especially the reduction of used angle pieces as well as the miniaturization of the vacuum size, increases the effective pump rate in the science chamber and hence improves the vacuum. For this, also the choice of using a glass cell is beneficial, as there is less hydrogen diffusing out of the material compared to a steel chamber. Details on the design can be found in Appendix~\ref{App: subsec: Vacuum design}. In this setup, we reach vacuum-limited trapping lifetimes above \SI{1000}{\second} in the 3D-MOT (for measurements see Appendix~\ref{App: subsec: Vacuum design}).

\begin{figure*}[t]
\includegraphics{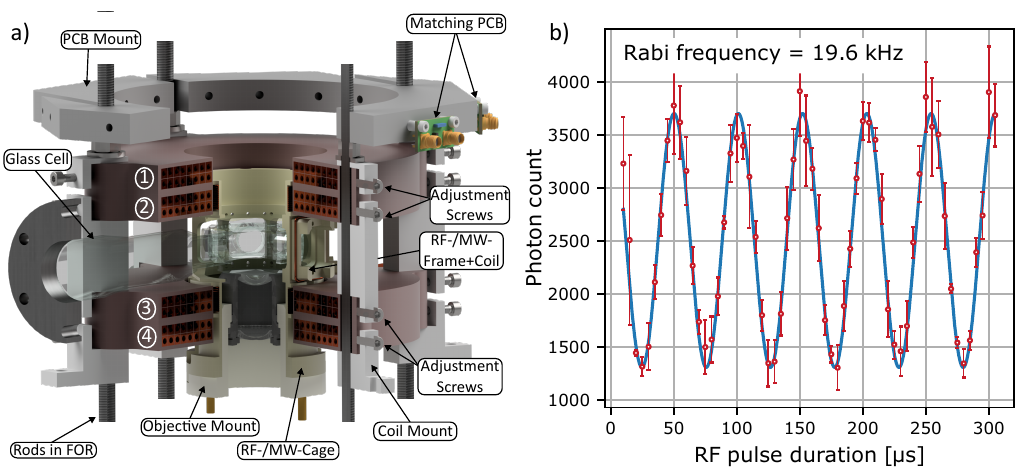}
\caption{\label{fig:coils}a)  A CAD rendering of the coil assembly in this configuration of the platform. For generating the DC offset fields we use four pancakes of 2\,x\,8 windings with 5\,x\,5\,mm$^2$ cross-section and \SI{3}{\milli\meter} diameter hollow core wires (labeled 1-4). These are positioned as close to the glass cell as possible to still fulfill Helmholtz configuration. Around the glass cell and the objective mounted inside the coils sits a CNC-machined polyetheretherketone (PEEK) cage \cite{RFcage} to which radiofrequency and microwave antennas can be mounted. These are connected to the matching circuits which are mounted above via a short SMA wire. b) RF transition Rabi rates. A typical Rabi oscillation curve for the $\ket{1} \rightarrow \ket{2}$ transition with rates of around \SI{20}{\kilo\hertz}.}
\end{figure*}

The miniaturization of the system and consequently the reduction of the distance between 2D- and 3D-MOT is also beneficial for achieving high 3D-MOT loading rates. This is because the 2D-MOT atom beam will have a residual transverse temperature and hence a beam divergence. Thus, for larger distances between the atom source and the science chamber, some atoms cannot be recaptured by the 3D-MOT reducing the effective loading rates. With this assembly, we realize 3D-MOT loading rates of \num{1.3d9} atoms/s, as measured by fitting a linear slope to the initial (non-saturated) loading in the 3D-MOT (see Figure~\ref{fig: MOT_chamber}b). The atom number is determined via fluorescence counting in a MOT (the calibration is detailed in Appendix~\ref{App: subsec: Atom number cal}).

To show the scaling possibility of this 2D-MOT configuration to even higher atom fluxes we measure the dependence of the loading rates into the 3D-MOT with the power in the 2D-MOT keeping the power in the 3D-MOT beams as well as the detunings the same. In the inset of Figure~\ref{fig: MOT_chamber}b one sees that the fluxes scale approximately linear with power showing that it should be possible to increase the loading rates of a $^6$Li 2D-MOT into the \num{e10}~atoms/s regime. This can be mainly done using more power, increasing beam waists and detuning, and increasing the oven temperature.

Also, it should be noted that the outer parts of the quadrupole field generated by the permanent magnets for the 2D-MOT can be used as the decreasing field in a Zeeman slower to decelerate the atoms coming from the oven. For this one only has to put an additional Zeeman slowing beam from opposite of the oven \cite{Lamporesi_2013}.


\subsection{\label{subsec:MagFields}Magnetic fields} 

In this section, we discuss the AC and DC magnetic field assemblies used in this configuration of the platform (shown in orange in Figure~\ref{fig:full-cad}). For $^6$Li, a broad Feshbach resonance around 800\,G is the main tool to control interactions between the atoms. Furthermore, the internal spin composition of the system, vital for effects like thermalization or pairing, can be precisely controlled by using radio-frequency (RF) or microwave (MW) fields. 

\subsubsection{\label{subsubsec:MagFieldsDC}DC magnetic fields} 

In Figure~\ref{fig:coils}a a CAD drawing of the magnetic field coils as well as the glass cell and the high-NA objective is given. The coils are placed in close proximity to the main high-NA objective which is the most demanding and most sensitive piece of optics in the experiment.

To ensure the long-term stability of the objective in terms of optical performance, and hence of the operative regions of all mPoCs, it is essential for the heat produced in the coils to be taken out efficiently, especially as the coils are positioned near the main objective. Hence, we prioritize efficient cooling over other design parameters like size or winding number, effectively realizing coils staying at room temperature at full duty cycle operation. For these coils the requirements are met by water cooling them via hollow-core wires, a technology proven to work in many experiments in the field of cold atoms. 

\begin{figure*}[t]
\includegraphics[width=1.00\textwidth]{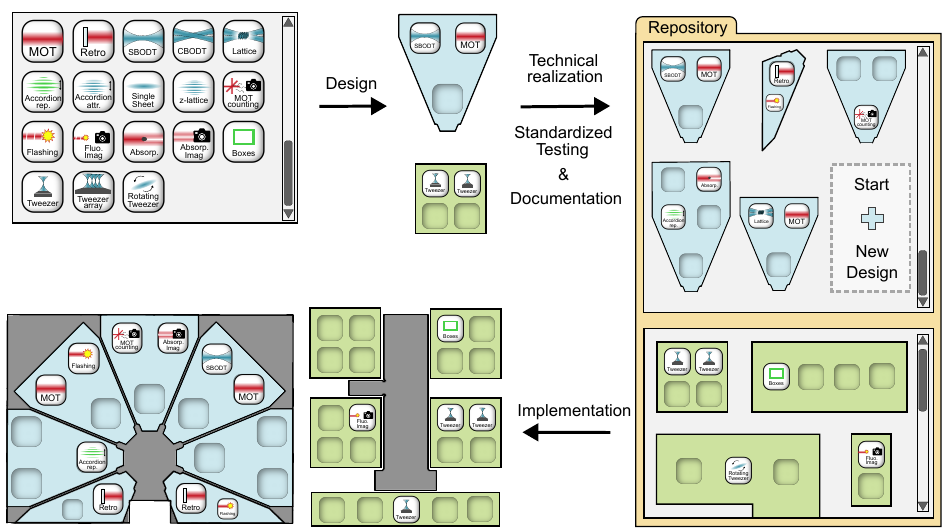}
\caption{\label{fig:Configuration}The typical workflow for configuring the modular platform, starting from an idea and going over design, realization, documentation and final implementation is shown. The workflow starts in the top left corner. Here, one can select from various functionalities (depicted as rectangular icons), which the PoC or mPoC should provide. All functionalities have internal DoFs, for example, a single beam ODT (SBODT) has as parameters the waist, trap frequencies, and depth of the required potential. This provides a clear framework for the design and the subsequent technical realization of the modules.With the standardized testing capabilities described in Appendix~\ref{App: subsec: PoCs}, it can be validated that the light fields meet the required performance.Afterwards, the design files are uploaded into a shared repository. In each laboratory, there is also a hardware repository of PoCs and mPoCs, which were built but are not needed in the experiment at the moment. The modules (PoCs and mPoCs) shown in the repository on the right-hand side are breadboards, which can differ in shape and also be enlarged while retaining modularity. The only requirement for this is that the precision holes are kept in the same position. Thus, the range of possible customizations and sizes of the modules is not limited by the modular design. Implementation into the experiment is straightforward as discussed in previous sections. There, one chooses existing modules from the repository to configure the platform, yielding a versatile range of functionalities. The final configuration shown in the lower left is the current configuration of our experiment.
}
\end{figure*}

Same as for all the optical modules also the positioning of the magnetic field assembly with respect to the frame of reference is critical. First, the position of the zero of the quadrupole field, generated by using the coils in anti-Helmholtz configuration, determines the position of a compressed MOT at a large magnetic field gradient. The vertical position of this center can be moved easily by tuning the offset current between the upper and lower coils, the horizontal alignment however has to be precise enough to be well inside the overlap of all ORs. As the position of the coils inside the resin cast is not specified to this precision, we add degrees of freedom to move the coils in the horizontal plane using adjustment screws shown in Figure~\ref{fig:coils}a.

Next, the magnetic field saddle of the offset magnetic field in Helmholtz configuration can differ from its geometric center due to the lead wires position or other imperfections to the radial symmetry. To move this saddle we will use tiny pieces of ferromagnetic steel to move the magnetic field lines and thus the saddle point. In this way, position accuracies of a few tens of microns were achieved in our group \cite{Lompe2011}.

With this assembly, we realize well positioned magnetic gradient and offset fields. The exact technical specifics are summarized in Appendix~\ref{App: subsec: DC Magnetic fields}.

\subsubsection{\label{subsubsec:MagFieldsAC}AC magnetic fields} 

For the control of the internal spin states, we use single-loop RF coils. These are mounted on a small cage made from Ultem inside the Feshbach coils at a distance of \SI{36}{\milli\meter} from the atoms. The impedance matching circuit of the antennas is placed above the coils and is connected via a short SMA cable to the antenna (see Figure~\ref{fig:coils}a). In this way, it is easy to connect the antenna in a modular way to different circuits that are matched to different frequencies (Appendix~\ref{App: subsec: RF coils}).

For the measurements (shown in Figure~\ref{fig:coils}b), the RF frequency is provided by a homebuilt DDS board and amplified using a \SI{100}{\watt} Minicircuits amplifier. With this setup, we realize Rabi rates of the $\ket{1} \rightarrow \ket{2}$ transition in $^6$Li of approximately \SI{20}{\kilo\hertz}. A typical measurement of this oscillation is shown in Figure~\ref{fig:coils}b. We attribute these large rates to our ability to place the antenna near the atoms as well as to not having conducting materials between antenna and atoms.


\section{\label{sec:Example}Platform configuration} 

A modularized platform accelerates and simplifies changes between configurations of an experiment with different functionalities. In Figure~\ref{fig:Configuration} this workflow is sketched out. Starting from a set of possible functionalities, which the PoC or mPoC is supposed to provide, and which are visualized as icons in the top left corner, a design for the optical setup is generated. This technical realization has properly consider the detailed properties of the required light fields such as size or depths of the resulting potentials. The construction and alignment is done on a test bench, where standardized testing of the light field distribution can be performed. In this way, it can be verified if the module meets the required functionalities.

These characterizations, including the size of the operative region, and the mechanical and optical design files can be documented and shared in a central repository. This repository can serve as a basis for close collaboration between different research laboratories on the design and exchange of optical hardware. Additionally, in each individual laboratory this repository can also be a storage of the actual hardware. Assembled modules, which are not needed in the experiment at the time, can be stored for later usage. In this way, the flexibility of the experiment increases with the number of assembled and tested modules. If there is no module with the desired functionalities in the repository, one can start a new design or modify an existing one. For this, the modularized platform offers an intuitive and easy-to-use framework for fast design and realization of new modules.

In the repository section of Figure~\ref{fig:Configuration}, we present some of the existing modules, designed among the two groups, which are right now setting up experiments within this platform, namely the UniRand Group in Munich \cite{Munich} and our group in Heidelberg. For the PoCs in the upper section, this includes modules for the generation of dipole traps with different geometries and wavelengths, addressing of atoms and optical cooling as well as fluorescence and absorption imaging. For the different mPoCs in the lower right corner, the modules contain tweezers at different wavelengths, waists, and trap frequency requirements as well as a DMD module for the generation of box potentials and a module for single atom- and spin-resolved free space imaging.  

Different combinations of PoCs and mPoCs then determine the functionalities of the experiments. In the lower-left corner of Figure~\ref{fig:Configuration} the current configuration for PoCs and mPoCs is presented. In this case, the functionalities include a MOT and optical dipole traps, imaging along different axes, tunable 2D confinement as well as repulsive optical box potentials, and high-resolution tweezers at different wavelengths.


\section{\label{sec:Outlook}Conclusion}

Modularization has shown to be a powerful design concept in a variety of fields \cite{Parnas72, Modularity_OneOff, Alexander1971, Maynard1972,Baldwin2000, Wiendahl2005}. We incorporated this concept of modularization into the design of quantum simulation experiments. With this, we benefit from similar improvements to those realized in other modularized systems. This opens up the possibility to combine the advantages of standardized technologies with the broad applicability of quantum simulation utilizing cold atoms in one versatile platform. The modular concept allows for increased flexibility of the setup, which yields robustness against future changes in requirements. On a day-to-day basis, this provides a framework in which upgrading of the optical hardware can be done whenever needed by implementing modules fitting the required functionality.

The functionality of each such module is independent of all other modules and can be tested individually using standardized testing methods, proving essential for time-efficient and precise verification of the output light fields. Furthermore, this modular concept opens up pathways towards standardization of such experiments and offers first working solutions. This shows up for example in the possibility of the PoCs to be used on all ports or in the standardized testing of modules.

For future developments, one could imagine that incorporating this concept can simplify the design and construction of increasingly complex platforms. Here, modularity helps to manage this complexity \cite{Baccarini1996, Baldwin2000} and to keep the optical setups disentangled. 

In addition, designing experiments in a modular way facilitates close collaboration of groups on a hardware level. In this case, an optical module developed by one group can easily be duplicated or adapted by all the others as designs can be found in a shared repository. Furthermore, not only optical hardware is shareable, but also mounting solutions connected to the frame of reference, such as the RF cage shown in Figure \ref{fig:coils}, developed by the UniRand team.

In a next step, one can also envision developing the modularity of the platform even further. This could include upgrades to the optical modules with, for example, motorized stages and the programming of an alignment procedure to enable the modules to align themselves. In combination with a suitable optimizer software, the implementation of a module could hence be fully automatized. To realize a fully plug-and-play platform one could also include the experimental control hard- and software in the modular concept.

There, one might envision a control software with an abstraction layer for "high-level programming" of the apparatus comparable to \cite{Qiskit}. Combined with the ability of fast reconfiguration of the platform this opens up a wide range of viable Hamiltonians with chooseable parameters, which end-users can select from.

With the modular \textit{Heidelberg Quantum Architecture} we contribute to the development towards on-demand programmable quantum simulation experiments, paving the way for their utilization as quantum technology platforms readily available for end-users.

\begin{acknowledgments}

We gratefully acknowledge many insightful discussions and close collaboration on the development of the presented apparatus with Naman Jain, Jin Zhang, and Maciej Gałka. 
Additionally, we gratefully acknowledge the support and fruitful discussions with the team of our mechanical workshop. In particular, we thank Denis Hoffmann, Alexander Worsch and Simon Rabenecker. 
We thank Fred Jendrzejewski for inspiring discussions during the initial planning phase of the vacuum design. 
This work has been supported by the Heidelberg Center for Quantum Dynamics, the DFG Collaborative Research Centre SFB 1225 (ISOQUANT), Germany’s Excellence Strategy EXC2181/1-390900948 (Heidelberg Excellence Cluster STRUCTURES), and the European Union’s Horizon 2020 research and innovation program under grant agreements No. 817482 (PASQuanS) and No. 725636 (ERC QuStA). This work has been partially financed by the Baden-Württemberg Stiftung. The work at MPQ received funding from the European Union’s Horizon 2020 research and innovation program under grant agreement No. 948240 (ERC UniRand). 

\textit{Author Contributions} T.H. and M.K. contributed equally to this work.

\textit{Competing Interest} The authors declare no competing interests.

\textit{Correspondence and requests for materials} should be addressed to T.H. (hammel@physi.uni-heidelberg.de) and M.K. (mkaiser@physi.uni-heidelberg.de).

\end{acknowledgments}

\bibliography{Main}

\begin{thebibliography}{40}
\bibitem [{CNC()}]{CNC}%
  \BibitemOpen
  \href@noop {} {}\bibinfo {note} {The PoCs, mPoCs and the frame of reference breadboard have been machined by a CNC mill Deckel FP-50-CC. The programmibilty is set to 1\,µm, however including wear of the drill as well as the positioning accuracy of the translation stages the actual measurable precision is in the range of about 10\,µm.}\BibitemShut {Stop}%
\bibitem [{Dic()}]{Dichroics}%
  \BibitemOpen
  \href@noop {} {}\bibinfo {note} {The dichroics are standard products by Semrock. $D_{1}$ (Di03-R561-t3) is a longpass dichroic mirror to reflect \SI{532}{\nano\meter} light out of the main beam path towards port $P1$. $D_{2}$ (Di03-R685-t3) is a longpass dichroic mirror to split out light with wavelengths $\lesssim\SI{698}{\nano\meter}$. This light is subdivided by a dichroic longpass mirror $D_{3}$ (Di03-R635-t3) into \SI{638}{\nano\meter} light for port $P2$ and \SI{671}{\nano\meter} light for port $P3$. Light transmitted by $D_{2}$ is split by a dichroic shortpass mirror $D_{4}$ (FF880-SDi01-t3) into \SI{1064}{\nano\meter} light for port $P4$ and \SI{813}{\nano\meter} light for port $P5$.}\BibitemShut {Stop}%
\bibitem [{RFc()}]{RFcage}%
  \BibitemOpen
  \href@noop {} {}\bibinfo {note} {The RF cage has been developed in the UniRand group, led by Philipp Preiss, and can be seen as one example for sharable hardware between experiments set up within this platform.}\BibitemShut {Stop}%
\bibitem [{Mun()}]{Munich}%
  \BibitemOpen
  \href@noop {} {}\bibinfo {note} {In the UniRand group the main contributors for the development of horizontal optical modules are Naman Jain and Jin Zhang.}\BibitemShut {Stop}%
\bibitem [{til()}]{tilt}%
  \BibitemOpen
  \href@noop {} {}\bibinfo {note} {The acceleration sensor used for alignment of the glass cell and the high-NA objective is the SCL3300-D01 by Murata.}\BibitemShut {Stop}%
\end{thebibliography}

\newpage

\appendix

\newpage
\,
\newpage

\section{\label{App: sec: Summary modules}Optical modules} 

\subsection{\label{App: subsec: Optics theory}Derivation of tolerances} 

In Section~\ref{subsec:modularity} the optical modules of this platform were presented. Precise and repeatable positioning of a module with respect to the atoms is a requirement for modularization. Here, we first derive the required tolerances for the modularization of optical setups from a geometrical optics viewpoint. We do this having the technical realization of the PoCs and mPoCs already in mind.

Starting with the tolerances of the PoCs, there are no fixed lenses or other optical components between the output of the PoC and the atom plane. The positioning of the PoC can differ from the optimal placement, in position and angle.

For a pure translational offset $s_{PoC}$ of the module compared to the defined position, the operative region shifts by the same amount $s_{OR}$.
\begin{equation*}
    s_{OR} = s_{PoC}
\end{equation*}

For a pure angular offset $\alpha$ around some axis of rotation within the module and distance $d$ of this axis to the atoms the operative region acquires a shift $s_{OR}$ of
\begin{equation*}
    s_{OR} = \tan(\alpha)\cdot d
\end{equation*}
and is tilted by the angle $\alpha$. Using a position tolerance of \SI{10}{\micro\meter} for both pins and a distance between both holes of $b$ = \SI{37.5}{\centi\meter}, this can maximally amount to a tilting of the operative region by \SI{53}{\micro\radian}. As the distance of the axis of rotation is approximately \SI{35}{\centi\meter} away from the origin of the frame of reference this can maximally amount to an offset of \SI{19}{\micro\meter}.

If there is a fixed lens inside the optical path and positional or angular shifts are with respect to this lens, these dependencies change. This is in particular the case for the mPoCs, where this lens is the high-NA objective. 
The lens focuses the light fields, effectively Fourier transforming them in the focal plane, and by this maps angle tolerances to positioning tolerances and vice versa. An angle $\alpha$ in front of the lens will generate a position shift $s_{OR}$ of the OR given by 
\begin{equation*}
    s_{OR}=\tan(\alpha)\cdot f
\end{equation*}
with $f$ being the focal length of the lens. Compared to the discussion earlier, here the angle maps on a pure translation in the focal plane.

In the same way a position shift by $d$ with respect to the optical axis of the lens results in an angle tilt $\alpha'$ of the OR of
\begin{equation*}
    \alpha' = \arctan\left(\frac{d}{f}\right).
\end{equation*}

When assuming having two lenses fixed in a telescope configuration and the module moving with respect to this telescope the dependencies are closely connected to the magnification of the telescope. 

There, a positional shift of $d'$ in front of the telescope results in a shift of
\begin{equation*}
    s_{OR} = \frac{d'}{M}
\end{equation*}
after the telescope, with the magnification $M = \frac{f_{2}}{f_{1}}$, with $f_1$ being the focal length of the lens close to the object plane and $f_2$ the focal length of the lens close to the image plane. Equivalently, an ingoing angle $\alpha'$ results in an angle 
\begin{equation*}
    \alpha = M\cdot\alpha'
\end{equation*}
of the light fields leaving the telescope.

The depth of the operative region $g$ in the atom plane scales with the square of the magnification $M^{2}$, when propagating it through the telescope.
\begin{equation*}
    g' = M^2\cdot g
\end{equation*}

In the realization of the mPoCs presented here, the high-NA objective, the relay telescope and the module lens realize a telescope with a magnification given by the choice of the module lens. The magnification is given by
\begin{align*}
    M(f_{ML}) &= \frac{f_1}{f_{obj}\cdot f_2}\cdot f_{ML}\\
    &= \frac{\SI{300}{\milli\meter}}{\SI{18.8}{\milli\meter} \cdot \SI{100}{\milli\meter}}\cdot f_{ML}[mm]\\
    &\approx \frac{1}{6}\cdot f_{ML}[mm].
\end{align*}

Hence, it is easy to realize magnifications in the order of 15 or higher by just placing a lens at a predefined position.

Having a fixed magnification setup between the module and the OR can loosen tolerance requirements for the positioning while increasing them for the angle when de-magnifying an intermediate image. However, the angle of the light field in the atom plane is often less critical for achieving a first signal compared to the absolute position. For example, loading into a tweezer requires overlap of the light field with the atoms, but the angle is uncritical. This angle can subsequently be optimized on the atoms doing trap frequency measurements, by characterization of the outgoing light field or as the position in an intermediate Fourier plane.

\subsection{\label{App: subsec: PoCs}Horizontal PoCs} 

\subsubsection*{PoC test bench}

\begin{figure}[t!]
\includegraphics{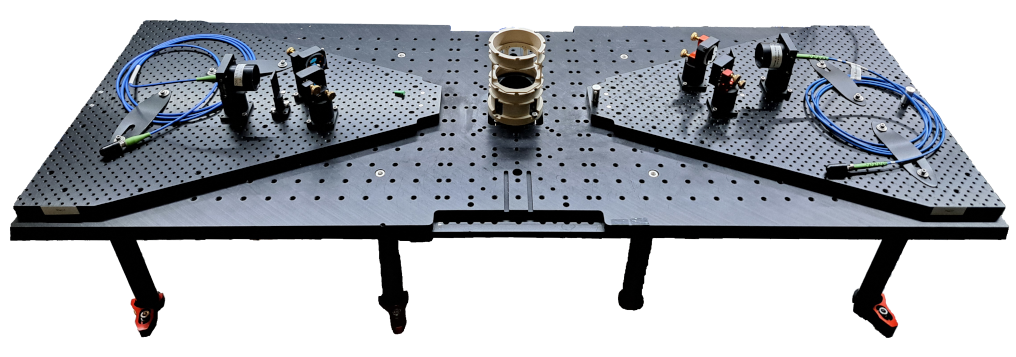}
\caption{\label{App: fig: testbench} The PoC test bench here set up with two PoCs and the PEEK frame to mount the insitu camera as done in the experiment and depicted in Figure~\ref{App: fig: insitu_cam}. In the up and down directions, in which no PoC is placed, two add-ons can be attached using the groves visible in the foreground and alignment pins, to provide a test bench for up to four PoCs.}
\end{figure}

The PoCs can be set up and pre-aligned using a test bench. The test bench, shown in Figure~\ref{App: fig: testbench}, is an identical copy of the interface for the PoCs on the frame of reference. A variety of sensors or mockups can be placed centrally on the test bench to enable development, construction, and most importantly testing and verification of the light field distribution of PoCs. 

This toolbox includes, but is not limited to, alignment targets, cameras and wavefront sensors. These sensors can be mounted on small diagnostic breadboards and be attached to the test bench using the same precision pin technique as the PoCs, making them easily exchangeable. In principle, up to four PoCs, at \ang{90} relative to each other, can be simultaneously placed on the test bench, in case beam travel between separate PoCs becomes necessary, e.g. for folded ODTs with beam recycling or optical lattices.

\subsubsection*{In-situ camera}
\begin{figure}[t!]
\includegraphics{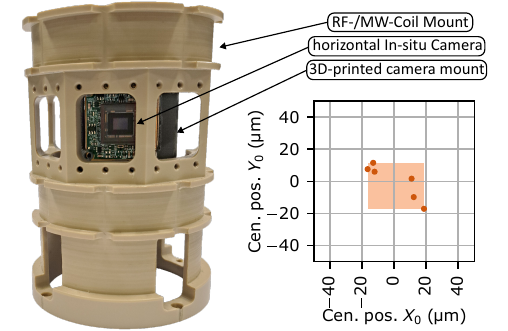}
\caption{\label{App: fig: insitu_cam} A copy of the PEEK-frame for RF- and MW coils with inserted side-ward facing in-situ camera for testing of PoC generated light fields. The inset shows repeatability measurements of placing the in-situ camera at the atom position yielding placement repeatabilities of $<\pm\SI{20}{\micro\meter}$. The orange box serves as a guide to the eye and has the minimum size to include the spread of all data points.}
\end{figure}

Furthermore, the light field distribution can not only be verified on the test bench but also in the experiment. As the vacuum is movable on a rail system a camera can be placed in the atom plane again measuring the light field distribution this time after implementation. Figure~\ref{App: fig: insitu_cam} shows the PEEK-frame for RF- and MW-coils introduced in Section~\ref{subsubsec:MagFieldsAC} to which the in-situ cameras can be mounted. On the left-hand side of Figure~\ref{App: fig: insitu_cam} the sideward pointing in-situ camera to optimize the PoCs when inserted into the PEEK holder is shown. As an in-situ camera, we use a FLIR machine vision board-level camera (FFY-U3-16S2M-C) and place it on a 3D printed mount that fits inside the PEEK-frame. The positioning is done using a mounting shoulder, serving as a stop for repeatable positioning of the camera chip in the center of the frame of reference. Additionally, the angular orientation is fixed by reference arms which lock the angle with notches near each each window to select the desired PoC Port. The inset in Figure~\ref{App: fig: insitu_cam} shows measurements of how repeatable the camera can be placed with this mount. For this measurement, a PoC with a collimated \SI{1}{\milli\meter} beam is sent onto the in-situ camera. Tracking the centroid of the beam when removing and re-inserting the camera quantifies the fluctuations of the camera positioning. We find placement repeatabilities of better than $\pm$\SI{20}{\micro\meter}.

Additionally, for the alignment of the mPoCs, a second downward-facing in-situ camera is used. As here positioning in the vertical direction is much more critical, due to the small Rayleigh length of the tweezers, we can only verify the relative positioning of the foci of traps. But, as the positioning of the focus in propagation direction maps onto the intermediate image plane scaled by the magnification squared, precise positioning of the focus on the mPoCs is sufficient to achieve an absolute position of the focus close to the origin of the frame of reference.

\subsubsection*{Positioning tolerances}
Passive alignment of optical modules requires substantial overlap of the operative regions (see Section~\ref{subsubsec: Optics theory}). Earlier, we derived how the desired properties of the ORs pose conditions on the required mechanical tolerances. In the following, we present and verify that this technical realization of the platform fulfills these requirements and comment on some challenges that need to be addressed.

For the horizontal PoCs, there is no additional optical setup connecting the module to the atom plane. This means that positioning and angle alignment errors of the PoC directly affect the operative region. Utilizing state-of-the-art CNC machining techniques with standard precision in the \SI{10}{\micro\meter} regime, the horizontal positioning capabilities are on the same order of magnitude.

\begin{figure}[h]
\includegraphics{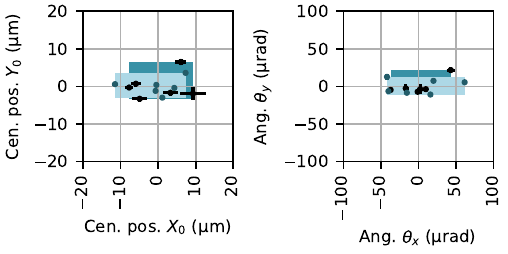}
\caption{\label{app: fig:repeatability} Repeatability measurements of placing two independent PoCs on the test bench using the pin-positioning mechanisms. Shown in the figure is the centroid position and angle of a collimated beam on the PoC w.r.t. a Shack Hartmann sensor in the atom plane when repeatably removing and placing the PoC on the test bench. Black and blue data points for centroid position and angle correspond to two independent PoCs, with their respective mean position and angle set as zero. Each data point is the mean of five wavefront measurements to account for external noise like background light. The errorbars indicate the corresponding standard deviation. The boxes are a guide to the eye to quantify the spread between the data points. The light blue box shows the spread of the blue data points, the dark blue box corresponds to the black datapoints. We find fluctuations of the centroid position of below $\pm$\SI{10}{\micro\meter} in horizontal (x-direction) and below $\pm$\SI{5}{\micro\meter} in vertical direction(y-direction). For the angle, we find fluctuations below $\pm$\SI{60}{\micro\radian} in the plane of the breadboard and below $\pm$\SI{15}{\micro\radian} perpendicular to it.}
\end{figure}

Both in-plane and vertical positioning repeatabilities with the pin mechanism have been measured and are presented in Figure~\ref{fig:modular_tech}b and \ref{app: fig:repeatability}. There, we placed a Shack-Hartmann wavefront sensor (Thorlabs WFS31-5C/M) in the atom plane on the test bench and illuminated it with a collimated \SI{3}{\milli\meter} diameter beam generated by a PoC positioned with its two pins. Using a wavefront sensor allows us to track the centroid position of the envelope as well as the angle of the beam when removing the PoC from the test bench and reinserting it again with the pins.

We observe centroid and angular fluctuations consistent with the expected tolerances of the mounting mechanism for two independent iterations of the measurement with independent PoCs, shown in Figure \ref{app: fig:repeatability} in different colors and observe consistent positioning capabilities across independently machined PoCs.

\begin{figure}[t]
\includegraphics{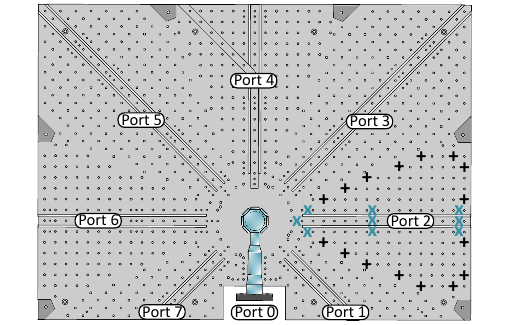}
\caption{\label{App: fig: PoC-Ports} Sketch of the mechanical frame of reference breadboard for the PoCs and the associated naming scheme of the ports. On port 2 the standardized interfacing points for the PoCs are shown. Blue markers (x) depict the precision holes for the pin mechanism, black markers (+) indicate through holes for M6 bolts to fasten the PoC to the frame of reference. Port 0 is the entrance port for the glass cell, but also features the common connection points, which can be utilized for placing in-situ sensors at the atoms position when the vacuum chamber is retracted.}
\end{figure}

The in-plane (x-direction) angular precision of the PoCs is limited by the added total clearance of the pins divided by their distance. Their baseline distance, depending on the chosen precision holes for referencing (marked in blue in Figure~\ref{App: fig: PoC-Ports}) can be up to \SI{375}{\milli\meter} yielding angle precision better than \SI{60}{\micro\radian}. Such an angular precision shifts the positioning of the OR by maximally \SI{19}{\micro\meter}.

The vertical (y-direction) angular precision is limited by the planarity of the PoCs and the optical breadboard or test bench acting as the frame of reference. To avoid bending of the frame of reference we manufactured a table-like monolithic aluminum (EN-AW5083) framework with a thickness of \SI{30}{\milli\meter}. The module breadboards are \SI{12.7}{\milli\meter} thick. The combined planarity yields in our case repeatabilities on the order of a few microns, see Figure~\ref{app: fig:repeatability}. 

We observe larger fluctuation in the plane of the breadboard than in the plane perpendicular to the breadboard. This means that the machined PoCs and test bench can be considered flat within the fluctuations due to the pin clearance. The dominant scale of positional fluctuations is set by the pin clearance. Positioning fluctuations along the insertion direction of the PoCs can be assumed to be similar to the x-direction fluctuations as both positions are constrained symmetrically by the pins. All fluctuations can be considered small compared to optical ORs for these low-NA Ports.

\begin{figure}[t]
\includegraphics[width=0.48\textwidth]{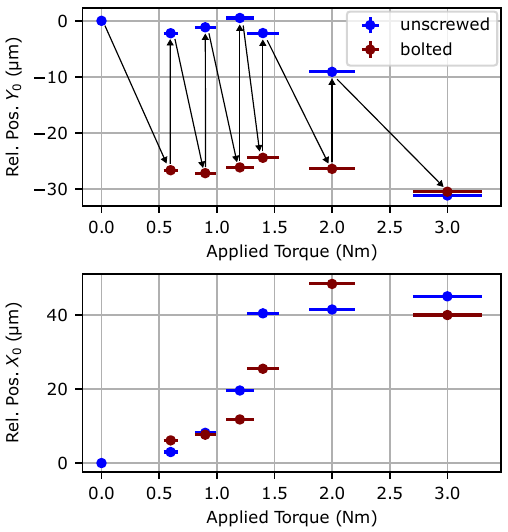}
\caption{\label{App: Fig: Torque} Drift of the centroid position of a beam from a PoC on the test bench observed on the in-situ camera in the atom plane depending on the applied torque on four M6 bolts with which the PoC was screwed down. X-direction corresponds to the plane of the PoC, Y-direction to the direction perpendicular to the PoCs. The upper plot shows the color coding for both plots and the order in which the measurements were taken.}
\end{figure}

Beyond positioning with pins, the PoCs can be fixed on the frame of reference or the test bench with screws for additional stability.
For this, the modules are screwed down from below onto the interface breadboard with M6 screws with a torque wrench for repeatability in mounting the modules. The precise way how to tighten the modules in place and the order in which the bolts are tightened are important for achieving optimal results. 

Figure~\ref{App: Fig: Torque} shows the results of a measurement to investigate the influence of the applied torque on the PoC position. We placed a PoC generating a collimated \SI{1}{\milli\meter} diameter beam on the test bench and gradually applied torque to the four M6 mounting screws and tracked the resulting beam centroid movement on the in-situ camera relative to the position only fixed by the pins. After tightening the bolts and measuring the beam position, we loosened the bolts to check for inelastic beam displacements during the mounting process.

In Figure~\ref{App: Fig: Torque}, one observes that for the y-direction, where the Poc lies flat on the test bench, two regions are observed. For lower torque, the bolts cause an elastic beam displacement, effectively pressing the PoC onto the test bench to cancel out small deviations in their flatness. When releasing the bolts, the beam position jumps back to its original pointing. Above a certain threshold of around \SI{2}{\newton\meter} in our case, the bolted position remains unchanged but the unbolted position does not jump back when releasing the bolts. This indicates that in this regime inelastic deformation starts to appear that deform the pointing of the optical setup. 

Shown in the lower plot in Figure~\ref{App: Fig: Torque}, in the x-direction one does not observe a switch between distinct positions when applying and releasing the bolts but a continuous deformation of the pointing with the applied torque, also after releasing the bolts. As this is the direction that is pre-locked by the pins, this indicates that the bolts potentially apply a shear force to the PoC leading to mechanical deformation. This deformation however seems to be small compared to the clearance of the pins for torques within the elastic regime of the y-direction. For now, this means that the applied torque should be limited to that regime when installing the PoCs.

The results of the most practical testing of the PoCs in shown in Figure~\ref{fig:modular_tech}c. This test mimics the complex everyday use of the PoCs and the decoupled development and debugging on a test bench separate from the experiment. Here, the defining quantity is how repeatable one can transfer a PoC from the test bench to the experiment. This transferability is dependent on multiple factors, among others, the effects of the tolerances of the pins, the precision holes, possible differences between the test bench and experiment in manufacturing, and the process of screwing the PoCs down.

\begin{table}[t]
\caption{\label{tab:PoCs} PoC summary for this configuration}
\begin{ruledtabular}
\begin{tabular}{p{1cm}p{7.6cm}}
Port & Designed light field\\
\hline
\\
\textbf{1} & \textbf{At \SI{671}{\nano\meter}}\\
& Retro reflection of a \SI{10}{\milli\meter} diameter MOT beam\\
& \textbf{At \SI{671}{\nano\meter}}\\
&Input of a resonant imaging beam beam (waist=\SI{600}{\micro\meter})\\
\\
\textbf{2} & \textbf{Unused}\\
& Empty port\\
\\
\textbf{3} & \textbf{At \SI{671}{\nano\meter}}\\
& MOT beam generation (waist=\SI{5}{\milli\meter}), centered, 0° angle with atom plane\\
& \textbf{At \SI{1064}{\nano\meter}} \\
& Horizontal ODT generation: waist=\SI{13.6}{\micro\meter}, centered, 0° angle with atom plane\\
\\
\textbf{4} & \textbf{At \SI{671}{\nano\meter}}\\
& Fluorescence imaging (Magnification 1, NA = 0.25)\\
\\
\textbf{5} & \textbf{At \SI{671}{\nano\meter}}\\
& MOT beam generation (waist=\SI{5}{\milli\meter}), centered, 0° angle with atom plane\\
& \textbf{At \SI{671}{\nano\meter}}\\
&Input of a flashing beam (waist=\SI{600}{\micro\meter})\\
\\
\textbf{6} & \textbf{At \SI{532}{\nano\meter}}\\
& Optical accordion, \SI{1.2}{\micro\meter} - \SI{23}{\micro\meter} lattice spacing, centered, 0°\\
\\
\textbf{7} & \textbf{At \SI{671}{\nano\meter}}\\
& Retro reflection of a \SI{5}{\milli\meter} diameter MOT beam\\
\end{tabular}
\end{ruledtabular}
\end{table}

To quantify this process, we placed in-situ cameras on the test bench and on the optical breadboard of the experiment. We use a PoC with an outgoing, collimated \SI{1}{\milli\meter}-sized beam. We placed it on the test bench, screwed it down with \SI{1.4}{\newton\meter} of torque on four M6 bolts and measured the position on the in-situ camera. Next, we unscrewed the PoC and repeated the mounting and position measurement on the optical breadboard of the experiment. We measure the distance between the centroid positions on both in-situ cameras and plot the deviation from their mean. This approach allows to cross out contributions from imperfect in-situ camera positioning (see Figure~\ref{App: fig: insitu_cam}).

We find that in this test the transfer of the PoC between test bench and experiment breadboard can be performed with repeatabilities of better than $\pm$\SI{30}{\micro\meter} in x-direction and $\pm$\SI{10}{\micro\meter} in y-direction. The observed behavior matches expectations based on adding up the tests of the pin tolerance and mounting-induced fluctuations with the bolts. In our case, the resulting positional variations are still sufficient to passively overlap the ORs of the PoCs with each other and with the ORs of the mPoCs.\\

As this provides a mechanical realization for the intended modularization of optical setups, we now want to use this modular character. In particular, we want to be able to place modules at different positions in the experiment without the need for reassembly.

The hole pattern in the breadboard forming the frame of reference, defining the connection point for the modules in this platform, is depicted in Figure~\ref{App: fig: PoC-Ports} and consists of nine precision holes for alignment and eighteen mounting holes. This is standardized for all seven PoC-Ports. This allows a PoC to be placed on all ports and setups can easily be swapped between ports.

In Table~\ref{tab:PoCs} the functionalities of the different PoCs on ports 1-7 in Figure~\ref{App: fig: PoC-Ports} are summarized. This corresponds to the configuration we are using for this experiment at the time this manuscript is written. This is also presented in the lower left corner of Figure~\ref{fig:Configuration}.

\subsection{\label{App: subsec: mPoCs}Vertical mPoCs} 


In Section~\ref{subsubsec:high NA}, multiple technical challenges have been discussed, which need to be addressed to modularize high-NA optics. Here, technical details are presented in the following order. First, the high-NA objective, its special design features, as well as the alignment of the vacuum viewport with the high-NA objective, is presented. Then, all mechanical mountings are illustrated including the objective, the relay telescope, and the distribution board. Here, we will elaborate on the gluing of dichroic mirrors. Lastly, we will elaborate on the optical designs of this platform including the relay telescope and possibilities for designs of the mPoCs.

\begin{table}[t]
\caption{\label{tab:SO objective} High-NA objective properties}
\begin{ruledtabular}
\begin{tabular}{lr}
\textbf{Parameters} & \textbf{Value}\\
\hline
&\\
\textbf{Numerical aperture} & 0.655 \\
\textbf{Working distance (total)} & \SI{13.7}{\milli\meter} \\
\textit{parts of this in}&\\
\textbf{Air} & \SI{1}{\milli\meter} \\
\textbf{Fused silica} & \SI{6.35}{\milli\meter} \\
\textbf{Vacuum} & \SI{6.35}{\milli\meter} \\
&\\
\textbf{Apochromatic at} & \SI{1064}{\nano\meter} \\
& \SI{671}{\nano\meter} \\
& \SI{532}{\nano\meter} \\
&\\
\textbf{Focal length} & \SI{18.8}{\milli\meter} \\
\textbf{Rear aperture} & \SI{24.8}{\milli\meter} \\
\textbf{Housing diameter} & \SI{45}{\milli\meter} \\
\textbf{Physical length} & \SI{40}{\milli\meter} \\
\textbf{Back focus distance} & \SI{1.9}{\milli\meter} \\
&\\
\textbf{Diffraction limited FOV}\\
\textbf{@ 671nm (measured)} & \SI{\pm 104}{\micro\meter} \\
\textbf{@ 532nm (theory)} & \SI{\pm 110}{\micro\meter} \\
\textbf{@ 1064nm (theory)} & \SI{\pm 130}{\micro\meter} \\
\textbf{Relative angle of optical} & (2.2 $\pm$ 1.6)\,mrad \\
\textbf{and mechanical axis} &\\
\end{tabular}
\end{ruledtabular}
\end{table}

The objective used is a custom-designed 0.655NA objective by Special optics. The design parameters and confirmed features are summarized in Table~\ref{tab:SO objective}. Here, we want to highlight the back focus distance of \SI{1.9}{\milli\meter}. This describes the fact that a collimated beam entering the objective from the atom side comes to focus \SI{1.9}{\milli\meter} outside of the housing. This enables us to place a $\lambda/4$-plate with a sputtered on point mirror with a diameter of \SI{1}{\milli\meter} in this plane. In this way, we can realize a vertical MOT beam going through the objective, getting back reflected into itself by the point mirror and thereby passing the $\lambda/4$-plate twice, acquiring the needed polarization shift. The small size of the point mirror only marginally influences the performance of diffraction-limited optics, like tweezers, as the area is small compared to the total area of the rear aperture. Making the waveplate dual-wavelength (e.g. $\lambda/2$ for 1064nm) further allows us for the realization of an infrared z-lattice in the same way.

For the vertical mPoCs, the operative region is limited by the FOV of the high-NA objective, which in our case is measured to be on the order of $\pm$\SI{100}{\micro\meter} \cite{Bunjes2022}. To actually achieve this measured size of the FOV in the experiment the relative angle of the vacuum window and the objective need to be perpendicularly aligned (for this objective to better than \SI{1}{mrad}), as otherwise aberrations are introduced into the optical path. This tolerance was simulated in Zemax using a blackbox file of the objective provided by Special Optics.

\begin{figure}[t]
\includegraphics{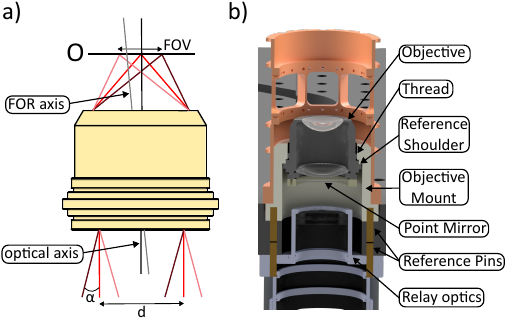}
\caption{\label{App: fig: objective_mounting} \textbf{a)} Sketch of the experiments high-NA objective indicating its focal plane with a diffraction-limited FOV of more \SI{\pm100}{\micro\meter} resulting in backside beams with a diameter $d$ of \SI{25}{\milli\meter} and an aperture exit angle $\alpha$ of \SI{\pm5}{\milli\radian} for the \SI{100}{\micro\meter} field-of-view. Important axes for the objective are the optical axis of its lens system, proven to be aligned well with its mechanical housing axis, and the central axis of the frame of reference which can in principle be both tilted and shifted with respect to the optical axis. \textbf{b)} CAD model of the objective mount in the experiment and the corresponding alignment mechanisms. The Peek-frame for RF- and MW-coils described in sec\ref{subsubsec:MagFieldsAC} is shown in orange encapsulating the objective mount.}
\end{figure}

To align this we tilt the miniaturized vacuum setup using spacers between the mounting plates connecting the rail system and the vacuum assembly (see Appendix~\ref{App: subsec: Vacuum design}). To measure the angle with respect to the frame of reference we use a tilt sensor \cite{tilt} with a resolution of \SI{100}{\micro\radian}, which can be placed on the glass cell and on the frame of reference. Minimizing the offset of the measured angles at these positions aligns the two components with respect to each other.

Care has to be taken as this only aligns the mechanical (housing) axis of the objective with the window, but not the actually important optical axis. It is hence important to verify that the optical axis of the objective overlaps with the mechanical axis of the objective housing. We measured the angle between the optical axis of the objective and its housing to be (2.2 $\pm$ 1.6)$\,$mrad \cite{Bunjes2022}. To make a decision if the angle between the window and optical axis has to be further adjusted one would need to measure this with higher precision. This we plan to confirm by performing trap frequency measurements in an optical tweezer at different positions within the operative region.

These measurements enable the passive positioning of the objective without any degrees of freedom within the frame of reference, which is advantageous for the modularization of the high-NA optical setups. The housing of the objective is screwed with its thread into a custom-made PEEK-mount having a well-defined stop for the objectives reference shoulder (see Figure~\ref{App: fig: objective_mounting}b) providing a mounting mechanism which is limited to the CNC-machining precision of the custom PEEK mount and the objective housing in the ballpark of tens of micrometer. 

The mount itself is positioned on the frame of reference using two precision holes and brass pins, such that by design the center of the frame of reference is well inside the operative region of the high-NA modules.

The mounting of the the relay stage, composed of lens tube components with an external support frame, is done via the same precision holes as the objective mount leading to the least possible misalignments between both. These tolerances are small compared to the rear aperture of the objective and the angle of the beams leaving the objective and hence for the passive mounting negligible.

The distribution optics (including dichroic mirrors and the module lenses) are mounted on a separate breadboard directly on the optical table with no direct connection point to the frame of reference. Indirectly, the distribution breadboard is referenced to the frame of reference via pins in the M6 holes of the optical table leading to translational positioning accuracies on the order of \SI{100}{\micro\meter}. Angle mismatch with respect to the frame of reference is limited by the combined flatness of the optical table and the distribution board. To avoid uneven deformations caused by the mounting screws, these were tightened using a torque wrench. The misalignments are however negligible compared to the collimated beam diameter of around \SI{8.3}{\milli\meter} and optical angles behind the relay stage of up to \SI{32}{\milli\radian}.

\begin{table}[t]
\caption{\label{tab:mPoCs} mPoC summary for this configuration}
\begin{ruledtabular}
\begin{tabular}{p{8.6cm}}
Port\\
Designed light field\\
\hline
\\
\textbf{P1 - 532\,nm}\\ 
DMD setup, configurable box potentials with a FOV of $\approx$\SI{200}{\micro\meter}\\
\\
\textbf{P2 - 638\,nm}\\
Unused - originally intended for near-detuned box potentials using incoherent light\\
\\
\textbf{P3 - 671\,nm}\\
Imaging, with a magnification of 2 using an Orca Quest v2 qCMOS camera\\
\\
\textbf{P4 - 1064\,nm}\\
Vertical ODT, waist=\SI{4.6}{\micro\meter}\\
Diffraction-limited tweezer, waist=\SI{0.51}{\micro\meter}, centered inside the FOV, 0° angle to atom plane\\
\\
\textbf{P5 - 813\,nm}\\
Single diffraction-limited tweezer, waist=\SI{0.39}{\micro\meter}, close to the $\ket{2p}\rightarrow\ket{3s}$ resonance in $^6$Li\\
\end{tabular}
\end{ruledtabular}
\end{table}

On this distribution board, the dichroic mirrors are placed and used to spectrally separate the mPoCs. Their reflectivity and transmission have been calculated to be operational for angle of incident variations of up to \ang{2} in our case. The relay telescope, introducing a de-magnification of 3, magnifies the angles by a factor of 3 up to a maximal angle of \ang{1.8} to still be inside the operative region. Hence, this works in our case but has to be taken into account when doing the Zemax simulations of the relay telescope as well as in designing the overall optical path between modules and objective and when choosing appropriate dichroic mirrors.

Furthermore, the dichroic mirrors also require extra care in the design of their mounts as uneven deformations of the glass plate introduce significant astigmatism. To minimize the stress of the glass while mounting the dichroic mirror, we took inspiration from \cite{gluepaper} and mounted them on a flat aluminum plate with four drops of evenly spaced silicone glue (Momentive TSE399C). This provides a fillet-bond-like mounting of the mirror with a flexible, slowly curing glue to minimize strain. The performance before and after gluing has been tested and is considered sufficient within the error margins of the measurement. 

Taking among other factors the physical size and the acceptance angles of the dichroic mirror into account, the exact lenses to use in the relay telescope have been found by extensive Zemax simulations using commercial-of-the-shelf lenses. The boundaries for optimization were the realization of an operative region for all required wavelengths, which should in principle be limited by the high-NA objective and not the relay telescope or the dichroic mirrors (including the option of custom-sized dichroic mirrors), as well as the physical size of the relay stage. In our configuration, the relay telescope had to be shorter than \SI{25}{\centi\meter} as it otherwise could not be mounted vertically behind the objective, which simplifies the passive mounting of these lenses with respect to each other and the frame of reference. Furthermore, the chromatic shift of the lens ensemble has to be taken into account and it has to be checked that for every required wavelength there is a port on the dichroic distribution board providing diffraction-limited performance for this wavelength.

By addressing these technical challenges we are able to modularize the high-NA optical setups in this platform. The configuration we chose for these modules is presented in Table~\ref{tab:mPoCs} and in the lower left part of Figure~\ref{fig:Configuration}.

\subsection{\label{subsec:Tools} A typical workflow} 

To exemplify the advantages the modularization brings to working with this platform we describe a real life example of implementing new modules into our machine.

We will step by step set up the system summarized in Table~\ref{tab:PoCs} and~\ref{tab:mPoCs}. The first challenge is to align a horizontal ODT with the compressed MOT, having a diameter of about \SI{100}{\micro\meter}. The cMOT has been approximately overlapped with the origin of the frame of reference by moving the coils as described in Section~\ref{subsec:MagFields}, determining the position using a camera. The horizontal ODT setup has been set up on the test bench and is implemented into the experiment using the pin mechanism. The light field is verified using the in-situ camera (see Figure~\ref{App: fig: insitu_cam}) and the ODT is aligned in focus and position to the center of this camera. In the first iteration without the alignment tools, the position of the focus in propagation direction was off by about 3\,-\,4\,mm. The alignment of this focus position is typically hard to optimize without access to the atom plane and was the critical parameter to correct using the in-situ cameras.

After removing the camera and inserting the glass cell again, we could immediately load atoms into the ODT. This procedure is reproducible so whenever the signal is lost it can easily be regained using the in-situ camera. On this first signal, optimization of the total atom number in the ODT has been performed by fine-tuning the positioning of the ODT within the cMOT. Getting the first signal took about 15 minutes from inserting the PoC to seeing atoms on the fluorescence camera.

This ODT we want to overlap with a vertical ODT along the objective axis. We prepare the mPoC by placing a camera in the correct intermediate image plane predicted by a Zemax simulation. We align position, angle, and focus with this camera and then place it in the experiment. At this point, we do not necessarily see a signal on the atoms. However, as we are certain that the operative regions of both modules overlap we now move the horizontal ODT along the horizontal direction until we see an atom signal in the vertical ODT on the camera. We move the horizontal ODT, because, as depicted in Figure~\ref{fig:Optics theory FOVs 2}, the ORs of the PoCs are typically much larger than those of the mPoCs, and hence the overlap region is often within the OR of the PoCs. Tuning only the horizontal degree of freedom of the horizontal ODT works because the Rayleigh length of the vertical ODT has a large enough extent in the propagation direction. Here also limitations of this alignment strategy become obvious, for example, directly overlapping a tweezer with a short Rayleigh length with a horizontal ODT with a small waist will still be a challenging task unless there are other traps, which can be used as guides as for example the vertical ODT shown here is a guide for positioning of the tweezer.

To add a tweezer into the system we upgraded the mPoC and added the optics producing a smaller spot in the intermediate image plane. We then overlap the tweezer and the vertical ODT on the camera in this plane in position and focus and directly see atoms loading into the tweezer. Fine alignment of relative positioning of horizontal and vertical ODT and the tweezer is then done on the atoms, optimizing again on the atom number.

Overlapping the light fields of additional mPoCs, such as tweezers at different wavelengths or optical box potentials (the current configuration of our experiment shown in Figure~\ref{fig:Configuration}) can be easily done by using the vertical in-situ camera. Placing this in the plane in which the tweezer is smallest (focused down to a single pixel of the camera) allows for alignment of position and focus of the light fields generated by the other mPoCs with respect to the tweezer.

\section{\label{{App: sec: MOT and Vacuum}}2D-MOT and vacuum}

\subsection{\label{App: subsec: Laser Distr.} Laser beams at 671nm} 

\begin{figure}[h]
\includegraphics[width=0.48\textwidth]{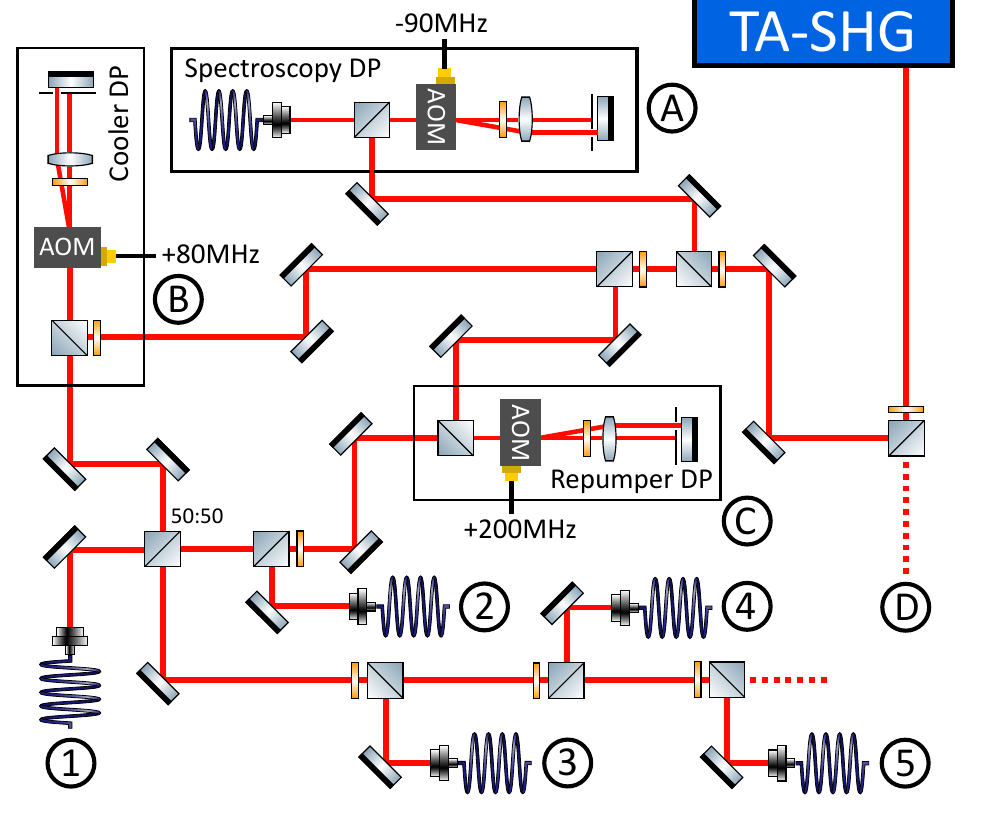}
\caption{\label{App: Fig: 671nm distribution} Schematic of the 671nm laser distribution. A: Spectroscopy double pass, B: Cooler double pass, C: Repumper double pass, D: Port for additional double passes. 1: 2D-MOT beam, 2: Push beam, 3-5: 3D-MOT beams.}
\end{figure}

For the platform described in Figure~\ref{fig:full-cad} a high flux atom source has been constructed. The favorable scaling of loading rates with optical power means that we need as much power as possible at the main cooling transition in $^6$Li, the $\ket{2S_{1/2}} \rightarrow \ket{2P_{3/2}}$ D2-line at \SI{671}{\nano\meter}. In recent years technical capabilities for high-power visible lasers increased tremendously yielding systems delivering several watts of power in a single mode, which is pivotal for increasing loading rates. Here, we use a Toptica TA-SHG with \SI{1}{\watt} of nominal output power. As now one such laser is in principle powerful enough to run more than one experiment, the question arises of how to set up the distribution optimally to support more than one experiment. 

Our design is sketched out in Figure~\ref{App: Fig: 671nm distribution}. There, we first split off some power directly after the laser head to use for locking with a homebuilt spectroscopy cell (A). For this, we pass the light through a double pass AOM setup, shifting the light by \SI{-180}{\mega\hertz} before using it to lock to the F=3/2 transition of the D2 line. This consequently locks the laser at \SI{-180}{\mega\hertz} detuned with respect to the cooler transition. To prepare the cooler and repumper light we then use a +\SI{80}{\mega\hertz} (B) and a +\SI{200}{\mega\hertz} (C) AOM double pass setup to shift the light back close to resonance. We can change the detuning from resonance up to about 10$\Gamma$ for cooler and 5$\Gamma$ for the repumper to the red-detuned side. As both frequencies are prepared in separate double pass setups the detuning of both can be adjusted individually and without realignment.

In Figure~\ref{App: Fig: 671nm distribution} following the frequency shifting optics the cooler and repumper beams are combined on a 50:50 beam splitter (ensuring the same polarization for both frequencies) and one arm is used for the 2D-MOT (1) and one arm is split up to the three 3D-MOT arms (2-4). A small portion of the repumper light is used for a push beam (5).

Shifting the frequency using double pass AOM setups is a pivotal step to be able to use one laser for multiple experiments as in this way the frequency of the laser itself is kept constant, making the light shareable between experiments. Given that there is enough power available to run two or more experiments with the same laser one can split out some light before going through the double passes (port D in Figure~\ref{App: Fig: 671nm distribution}) and build two additional paths for cooler and repumper of a second experiment. At the time this manuscript is written, we are using all the laser power for the experiment presented here, so no power is coupled into port D.

\subsection{\label{App: subsec: Vacuum design} Vacuum design}

\begin{figure}[htb]
\includegraphics[width=0.48\textwidth]{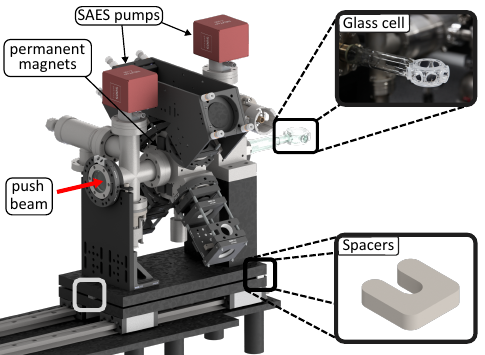}
\caption{\label{App: Fig: 2D-MOT tilting} A CAD rendering of the vacuum assembly including a 2D-MOT, Ion-Getter pumps in HV and UHV region as well as the glass cell science chamber (top inset) and the push beam. The position of the horseshoe spacers, which enable precise tilting of the whole assembly is indicated by small rectangles, and the shape of the spacers is depicted in the bottom inset.}
\end{figure} 

Here, we present the miniaturized vacuum design used in this configuration of the experimental platform. The vacuum is separated into two parts: A high vacuum (HV) region, which the oven and the 2D-MOT chamber are part of, and an ultra-high vacuum (UHV) region consisting mainly of the science cell.

To ensure an ultra-high vacuum in the science chamber the HV and UHV region are disentangled by a differential pumping stage, in our case a \SI{44.5}{\milli\meter} long and \SI{2}{\milli\meter} wide conical tube (opening angle \SI{10}{\milli\radian}). Theoretically, this tube has a conductance of approximately \SI{0.15}{l/\second} in the molecular flow regime.

On the UHV side, we chose a small octagonal glass cell as the main science chamber for its high numerical optical access and the possibility of placing magnetic field coils and antennas close to the atoms (see Section~\ref{subsec:MagFields}). 

The glass cell used is manufactured by Precision Glassblowing. The side window thickness is customized to \SI{3.5}{\milli\meter} to enable the use of commercially off-the-shelf G-Plan Mitutoyo objectives with a NA of 0.3. The coating of all windows on both in- and outside is an anti-reflection nano-texture coating by Telaztec (RAR.L2).

In Figure~\ref{App: Fig: 2D-MOT tilting} a picture of this octagonal glass cell is shown. Already here the very low reflectivity of the viewport surfaces up to large angles can be seen by comparing the reflections on the viewports with the reflections on the glass cell neck.

The oven and 2D-MOT section of the vacuum is connected via a standard four-way cross to a SAES Z100 Ion-Getter pump (left red pump in Figure~\ref{App: Fig: 2D-MOT tilting}). The science chamber section is connected via a custom-made connector to a SAES Z200 pump (right red pump in Figure~\ref{App: Fig: 2D-MOT tilting}).

The vacuum assembly is mounted onto a base plate, which is connected to a second base plate by horseshoe-shaped spacers (shown in the inset of Figure~\ref{App: Fig: 2D-MOT tilting}). Changing the relative height of these spacers enables tilting of the vacuum setup in both spatial directions. The spacers have been machined with height differences in steps of \SI{10}{\micro\meter} resulting in an angle resolution for tilting of around \SI{100}{\micro\radian}. This is sufficient resolution for fine alignment of the glass cell window with respect to the high-NA objective.

An integral parameter of any cold atom experiment is the vacuum-limited lifetime given by collisions with background gas atoms. This can show up in various different ways in the final performance of the machine for example in the fidelity of preparing a specific state or in the temperature and particle number of a sample.

Due to the miniaturized and simplified vacuum design described before we expect the lifetimes in this machine to be on the order of several minutes, comparable to or better than already running machines in our group.

\begin{figure}[t]
\includegraphics[width=0.48\textwidth]{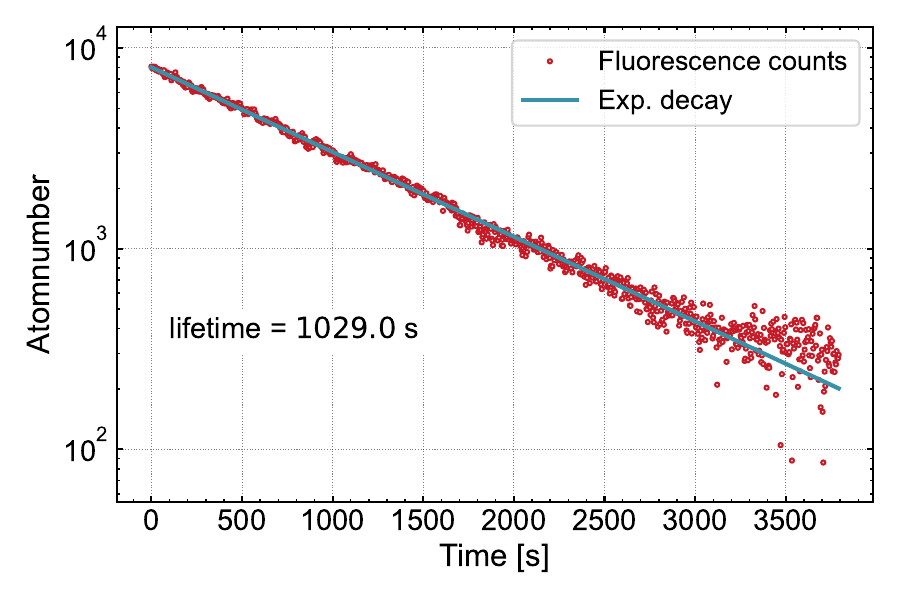}
\caption{\label{App: fig: Vacuum lifetime} Vacuum limited decay curve of a sample of initially a few thousand atoms captured in a 3D-MOT. The fit is done with an exponential decay function. The characteristic timescale, the vacuum limited lifetime, is determined to be around \SI{1030}{\second}.}
\end{figure}

To measure this, we prepare a 3D-MOT of a few thousand atoms and measure the fluorescence over an extended period of time. We see that the fluorescence decays slowly, which we attribute to atoms colliding with background gas atoms and getting lost from the trap. By fitting an exponential decay we can extract the characteristic timescale of this process, which is the vacuum limited lifetime. A typical decay curve measured in this experiment is shown in Figure~\ref{App: fig: Vacuum lifetime}. This type of measurement we repeated three times in intervals of about one year starting in March 2023 to check how the vacuum lifetime evolves. We measure an increasing trend for the lifetimes rising from \SI{500}{\second} (March 23) to \SI{970}{\second} (May 24) and to \SI{1030}{\second} (December 24).

\subsection{\label{App: subsec: Atom number cal}Atom number calibration} 

A central tool for the quantitative characterization of the capabilities of this platform is the determination of atom numbers. This is for example essential to quantify the loading rates of our 2D-MOT atom source. For this atom counting we use a standard approach in which we collect the fluorescence light the atoms spontaneously emit into the full solid angle when they fall back into the groundstate, while cycling on the D2 line transition in the 3D-MOT. We use a 1", f\,=\,\SI{50}{\milli\meter} lens to collect photons at a large solid angle and focus this on a machine vision camera. By integrating the signal in a defined region of interest and knowing the scattering rates, detection efficiencies, and solid angles of the imaging setup, we can calculate back from this signal how many atoms this corresponds to \cite{Hume2013}.

\begin{figure}[b]
\includegraphics[width=0.48\textwidth]{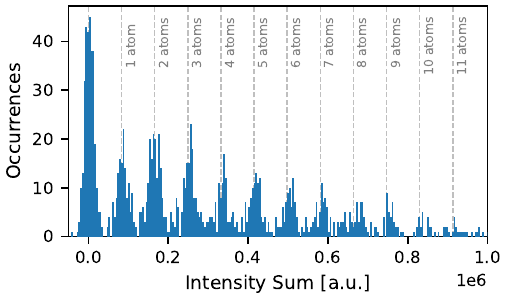}
\caption{\label{App: fig: counting_cal} Atom number calibration. When integrating the fluorescence signal of a MOT for a few atoms inside a small region of interest and plotting the number of occurrences of a particular signal strengths in a histogram individual atom number peaks emerge. Here, we present a typical data set of our machine for such a measurement. From left to right the peaks correspond to the background, i.e. zero atoms, then one atom, two atoms, etc. We use the single atom intensity, which is given by the distance between the zero and one atom peak so in this case about 80.000, to calibrate the atom numbers in this experiment.}
\end{figure}

Additionally, by lowering the background as much as possible it is possible to distinguish individual atoms by their amount of fluorescence. In this way, clear atom number peaks are visible up about ten atoms in our case (see Figure~\ref{App: fig: counting_cal}). The average fluorescence emitted by one atom can then be used to determine the atom number for larger systems.

To also calibrate the atom numbers at large numbers of more the $10^7$ we cannot use our counting MOT anymore as this cannot hold so many atoms. As we need to use different MOT parameters (especially further detuned cooler and repumper frequency) we need to recalibrate our counting for larger samples. For this, we use a sample of about $10^4$ atoms (measured with the counting MOT) and measure the collected signal strength from a MOT at parameters at which it can hold more than $10^9$ atoms. We measure that the fluorescence per atom, when using MOT parameters suitable for large atom numbers, is reduced by approximately 30\%. The loading rates presented later are taken by dividing the total fluorescence signal by the single atom signal in the counting MOT. Hence, we slightly underestimate the number of atoms of large samples by approximately 30\%.

\subsection{\label{App: subsec: Oven}Oven and 2D-MOT chamber design}

\begin{figure}[h]
\includegraphics[width=0.48\textwidth]{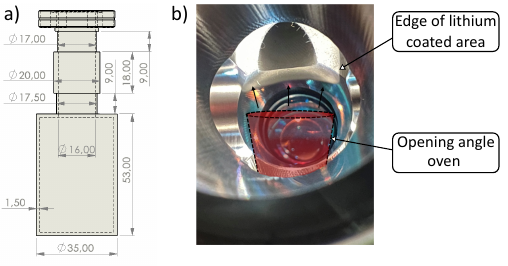}
\caption{\label{App: fig: Oven} a) The oven design \cite{Hammel2021} for this configuration of the platform. The dimensions of the reduced wall thickness segments are calculated assuming the 2D-MOT chamber as a heat bath at room temperature. We put two such regions to make sure that the neck remains above melting temperature up to the second narrow segment to reduce the possibility of clogging. b) Photo of the inside of the chamber along the in-coupling viewport of the 2D-MOT. The theoretical opening angle of the oven is shown in red and at the top part of the chamber the accumulated lithium can be seen as a white surface.}
\end{figure}

To scale up loading rates a suitable oven design is needed \cite{Tiecke_2009,Hammel2021} providing large initial fluxes while being able to run for long times such that it does not have to be replaced regularly. We use a custom-made oven with an aperture radius of r\,=\,\SI{8}{\milli\meter} (see Figure~\ref{App: fig: Oven}), which is filled with \SI{9.0}{\gram} of enriched $^6$Li and \SI{0.1}{\gram} of lithium in natural abundance, which is predicted to run for more than 10 years at a standard oven temperature of about \SI{350}{\degreeCelsius}. At this temperature, we predict atom fluxes coming from the oven of approximately $1\times 10^{16}\, \text{s}^{-1}$.

In this design, the oven is attached to the custom 2D-MOT chamber (see Figure~\ref{fig: MOT_chamber}a) at the bottom CF16 flange. The oven is custom made and the length of the neck of the oven tube has been adjusted to not have a direct line of sight with the CF40 viewports used to shine in the 2D-MOT cooling beams, preventing hot lithium atoms which are not captured by the 2D-MOT from coating the vacuum viewports. When looking into the chamber through one of the viewports of the 2D-MOT chamber near the oven, one can see where the lithium coats the walls of the chamber. This is shown in Figure~\ref{App: fig: Oven}b and matches our expectations. 

Along the neck of the oven, there are two parts with reduced wall thickness acting as heat blocks to thermally disconnect the oven from the chamber and the attached permanent magnets.

An additional viewport along the 2D-MOT axis is used for a push beam to steer the atoms through the differential pumping stage. We found this push beam and its properties (detuning, power, alignment) to be highly influential for the final loading rates.

This miniaturized chamber has predefined places for four stacks of neodymium permanent magnets generating the quadrupole field needed for the 2D-MOT \cite{Lamporesi_2013}. For the magnets custom holder attach them to the vacuum chamber, with which they can be positioned with a precision of \SI{1}{\milli\meter}. In our case, we found the positioning of the magnets to be optimal when they are placed at the theoretically optimal position, centered symmetrically around the 2D-MOT.

Also, for the setup described here, the 2D-MOT has several advantages over a Zeeman slower. Most importantly it reduces the amount of hot lithium atoms and dirt from the oven reaching the science chamber. Further advantages include the smaller size of the vacuum setup and the possibility to operate the cold atom source using permanent magnets. Both facilitate miniaturization and movability on a translation stage.

Next, as we aim in our configuration of the platform for sub-second cycle times of the experiment, this poses a challenge for moving mechanical components in terms of speed and longevity. In particular, for a Zeeman slower this can be challenging, as a mechanical shutter is required to shut off the atomic beam.

Hence, as long as the 2D-MOT can yield high loading rates it is favorable for meeting the goals of this platform.

\subsection{\label{App: subsec: 2D-MOT performance}2D-MOT performance} 

\begin{figure}[t]
\includegraphics[width=0.5\textwidth]{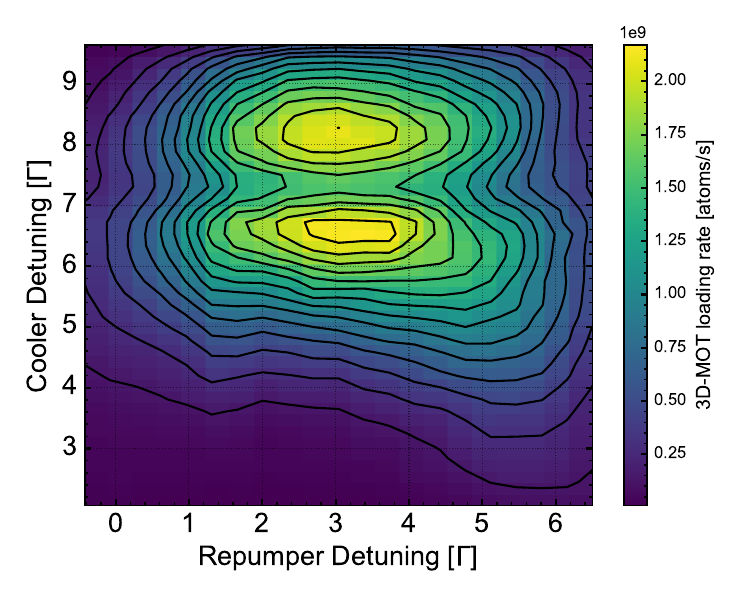}
\caption{\label{App: fig: detuning loading rate} Dependence of the 3D-MOT loading rates, depicted in false colors, on the cooler and repumper detuning in 2D- and 3D-MOT.}
\end{figure}

This modular platform is usable for a wide range of atomic species, for many of which a 2D-MOT as a cold atom source has already been shown to yield reasonable loading rates \cite{Chomaz23,Jend22,Nosske_2017,Dorscher_2013,Lamporesi_2013,Tiecke_2009}. Here, we will give a brief characterization of the most important tuning parameters of the 2D-MOT presented in Section~\ref{subsec:Vacuum}.

In this setup, a single beam 2D-MOT in a bow-tie configuration is set up with the goal to scale up loading rates. The beam is generated by directly collimating the light coming from a NA\,=\,0.12 polarization maintaining, single mode fiber (Thorlabs P3-630PM-FC-10) using a f\,=\,\SI{150}{\milli\meter}, 2” Achromat (Thorlabs AC508-150-A-ML), which results in a beam diameter of about \SI{36}{\milli\meter}. It has been seen that using smaller beams can increase the sensitivity to alignment and power balancing as larger parts of the low-intensity tails are seen by the atoms, causing the 2D-MOT to bend. This makes it difficult to steer the atoms through the differential pumping tube, effectively reducing the total flux arriving at the 3D-MOT.

Given the optical distribution of the \SI{671}{\nano\meter} light (see Appendix~\ref{App: subsec: Laser Distr.}) the detuning of the cooler and repumper light is the same in 2D- and 3D-MOT. In Figure~\ref{App: fig: detuning loading rate} this detuning was optimized yielding optimal loading rates at $\Delta_{rep}$ = $3\Gamma$ and $\Delta_{cool}$ = $8\Gamma$. As in previous works on $^6$Li 2D-MOTs \cite{Tiecke_2009} two peaks are visible where in our case both peaks yield very similar loading rates.

\begin{figure}[b]
\includegraphics[width=0.48\textwidth]{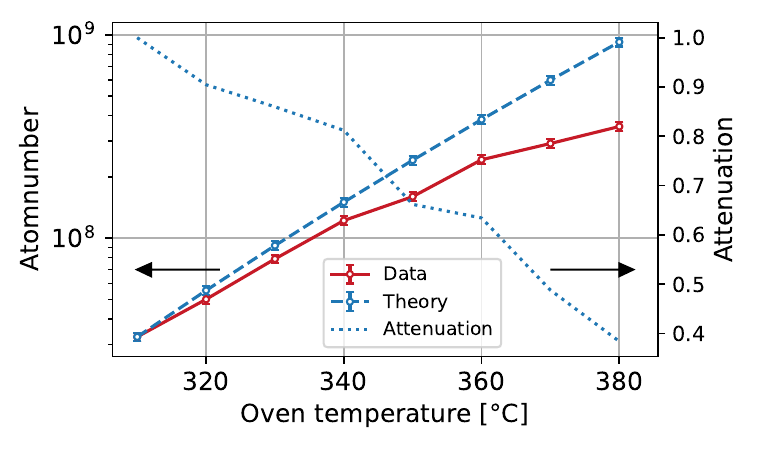}
\caption{\label{App: fig: oven temp loading rates} Dependence of the 3D-MOT loading rates (shown in red) on the oven temperature. For the theory line (dashed blue) we scale up the loading rate at \SI{310}{\degreeCelsius} with the expected scaling of the oven flux with temperature \cite{Hammel2021}. At larger temperatures there is an attenuation of the measured loading rates compared to the expected ones. This is quantified as the ratio of both values, and is shown as the blue dotted line.}
\end{figure}

For this measurement, we used the full power we had available, which corresponds to \SI{153}{\milli\watt} (cooler) + \SI{44}{\milli\watt} (repumper) in the 2D-MOT and \SI{50}{\milli\watt} (cooler) + \SI{14}{\milli\watt} (repumper) in each 3D-MOT beam. For the 3D-MOT we use beams with waists of \SI{5}{\milli\meter}, hence operating in the saturated regime ($I_{sat} = 2.54\frac{mW}{cm^2}$) for the chosen beam sizes of 2D- and 3D-MOT.

All measurements shown in this manuscript are taken at an oven temperature of \SI{350}{\degreeCelsius}. From the temperature dependence of the saturated vapor pressure of $^6$Li \cite{Alcock1984}, one would expect an increase of flux with temperature by about an order of magnitude every \SI{50}{\kelvin}. To verify this, we measure the 3D-MOT loading rate in the regime from \SI{310}{\degreeCelsius} to \SI{380}{\degreeCelsius}. This is shown in Figure~\ref{App: fig: oven temp loading rates}. This data has been taken at a non-optimal configuration of the 2D-MOT as there was misalignment in the laser distribution and thus non-optimized values for different parameters like cooler/repumper balancing at the time. This also shows that achieving loading rates of \num{1.3e9} atoms/s is intricately coupled to a stable laser distribution and beam alignments of 2D- and 3D-MOT.

\begin{table}[t]
\caption{\label{tab:2D-MOT parameters} 2D-MOT and 3D-MOT parameters}
\begin{ruledtabular}
\begin{tabular}{lrr}
\textbf{Parameter} & \textbf{Value} & \textbf{Unit}\\
\hline
&&\\
\textbf{2D-MOT}& &\\
\hline
&&\\
\textbf{Beam waist} & 15.6 & mm\\
\textbf{Power (Cooler)} & 120 & mW\\
& 6.2 & $I_{sat}$\\
\textbf{Power (Repumper)} & 80 & mW\\
& 4.1 & $I_{sat}$\\
\textbf{Detuning (Cooler)} & 8 & $\Gamma$\\
\textbf{Detuning (Repumper)} & 3 & $\Gamma$\\
&&\\
\hline
&&\\
\textbf{Push beam} &  &\\
\hline
&&\\
\textbf{Beam waist} & 1 & mm\\
\textbf{Power} & 340 & \si{\micro\watt}\\
\textbf{Detuning} & 3 & $\Gamma$\\
&&\\
\hline
&&\\
\textbf{3D-MOT} &  &\\
\hline
&&\\
\textbf{Beam waist} & 5 & mm\\
\textbf{Power (Cooler)} & 40 & mW\\
& 20 & $I_{sat}$\\
\textbf{Power (Repumper)} & 20 & mW\\
& 10 & $I_{sat}$\\
\textbf{Detuning (Cooler)} & 8 & $\Gamma$\\
\textbf{Detuning (Repumper)} & 3 & $\Gamma$\\
\end{tabular}
\end{ruledtabular}
\end{table}

In Figure~\ref{App: fig: oven temp loading rates} one can see that the loading rates at larger oven temperature are attenuated with respect to the expected flux, when just scaling up the flux at \SI{310}{\degreeCelsius} with the known increase in flux from the oven. This behaviour is expected and discussed in detail in \cite{Tiecke_2009}. There, the attenuation seems to set in at slightly larger oven temperatures, which could be attributed to the non-optimized parameters in our case and the different oven and 2D- and 3D-MOT geometry.

The optimal parameters for this high-flux atom source are summarized in Table~\ref{tab:2D-MOT parameters}. The oven temperature is again set to \SI{350}{\degreeCelsius}.

\section{\label{App: sec: Magnetic fields}Magnetic fields} 

\subsection{\label{App: subsec: DC Magnetic fields}DC Magnetic fields} 

For the generation of DC offset fields as well as gradient fields in this experiment, we use four coils with 2\,x\,8 windings each (see Figure~\ref{fig:coils}a). The small number of windings is a compromise between using large cross-section wires (5\,x\,5\,mm$^2$ with a \SI{3}{\milli\meter} diameter hollow-core) and keeping the coils small and hence close to the atoms. The large cross-section is beneficial as it allows for larger hollow cores, which support more efficient cooling. For this cooling of the coils, we use a \SI{16}{\bar} pump. This enables us to take out large amounts of heat from the coils. Hence, we can let them continuously run at \SI{440}{\ampere}, corresponding to magnetic offset fields of \SI{>2000}{G}, resulting in steady-state temperatures of the coils of about \SI{30}{\degreeCelsius}. As the duty cycle of the coils is much lower than this and we typically need much less current we found a typical steady state temperature of the coils during operation to be 1-2\,\SI{}{\degreeCelsius} above room temperature. By adjusting the temperature of the cooling water we predict that for a wide range of duty cycles we can operate the coils exactly at room temperature.

For the magnetic field configurations used in this experiment, it is essential to control the offset, gradient, and curvature of the field. In technical terms, this means that different combinations of currents $I_i$ flowing through the coils, shown in Figure~\ref{fig:coils}a, need to be stabilized independently. The total current ($I_1+I_2+I_3+I_4$) determines the offset field, the current difference between upper and lower coils ($(I_1+I_2) - (I_3+I_4)$) determines the gradient, while the difference between inner and outer coils ($(I_1+I_4) - (I_2+I_3)$) determines the curvature. 

To stabilize these three quantities we use for the four coils three individual power supplies with the inner coil pair (2+3) being connected in series. The currents in these three circuits are non-invasively measured individually using current transducers (LEM IN-500-S) and measuring the voltage drop along a precision resistor. This is then used to feedback on the power supplies using a digital PID loop to stabilize the current. 

As in this setup, the inner coil pair is located closer to the atoms than Helmholtz configuration this pair will add a radially anti-confining harmonic term to the offset fields, while the outer pair will add a confining potential. In particular, by changing the current difference in the outer and inner coil pair, the radial potential can be tuned from confining to anti-confining opening up interesting possibilities for matter-wave magnification protocols and the generation of large homogeneous box traps. 

As we often want to quench magnetic fields as fast as possible the coils can be switched on and off using assemblies of ten MOSFETs (IRFB3077PbF) placed in parallel. In this way, the large offset magnetic fields can be switched with a time constant of \SI{2000}{G/\milli\second}.

\subsection{\label{App: subsec: RF coils}RF coils and matching circuits} 

The RF coils presented in Section~\ref{subsubsec:MagFieldsAC} are single-loop coils wound around a holder to connect to the RF cage. The coil has dimensions of \qtyproduct{30 x 22}{\milli\metre} and a distance of \SI{36}{\milli\meter} to the atoms. Via a short SMA cable, the coils are connected to a matching board on which their impedance is matched to \SI{50}{\ohm}. On the small PCB board, a capacitor in series ($C_s$) and one in parallel ($C_p$) connects the coil to the amplifier. Choosing both capacities correctly one can match the impedance to \SI{50}{\ohm} at the desired frequency, in our case the transition frequencies of the $\ket{1} \rightarrow \ket{2}$ and $\ket{2} \rightarrow \ket{3}$ transition in $^6$Li.

In this setup we can drive both transitions with the same coil and matching board, but with different Rabi frequencies. For the originally designed frequency (\SI{76.5}{\mega\hertz}) we measure Rabi frequencies of \SI{20}{\kilo\hertz}, while for the non-perfectly matched frequency (\SI{84.2}{\mega\hertz}) we measured Rabi frequencies of about \SI{6}{\kilo\hertz}. This shows that the antennas have a large tuning range, with the drawback that at the non-matched frequency the Rabi rates are reduced.

To avoid this, we can use a second coil with its own matching circuit optimized at \SI{84.2}{\mega\hertz}. Due to the modular structure, we can place multiple coils around the cell. This opens up interesting possibilities for the engineering of more complex RF fields using multiple antennas.

\begin{figure}[h]
\includegraphics[width=0.48\textwidth]{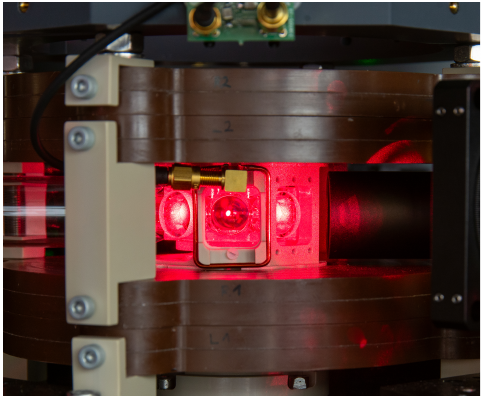}
\caption{\label{App: Fig: MOT photo} A picture of the surroundings of the glass cell including the RF-cage with an attached antenna and its matching circuit above. In the center of the glass cell the atoms in a MOT are visible.}
\end{figure}

In Figure~\ref{App: Fig: MOT photo} a photo of the heart of the experiment is presented, showing the glass cell with the high-NA objective below, the Feshbach coils, an RF coil, and most importantly the atoms in a MOT sitting at the origin of the frame of reference of this modular platform.

\FloatBarrier

\end{document}